
\magnification=1200
\noindent
\font\klein=cmr8 scaled\magstep0
\font\gro=cmr12 scaled\magstep2
\font\mit=cmr8 scaled\magstep2
\def\crns{\cr\noalign{\smallskip}}
\def\crnm{\cr\noalign{\medskip}}

\def\q#1{\lbrack #1 \rbrack}
\def\smno{\smallskip\noindent}
\def\meno{\medskip\noindent}
\def\bigno{\bigskip\noindent}
\def\pano{\par\noindent}
\def\oG{\overline{G}}
\def\o#1{\overline{#1}}

\def\Dt{\Delta}
\def\la{\langle}
\def\ra{\rangle}
\def\lb{\lbrack}
\def\rb{\rbrack}
\def\ds{\displaystyle}
\def\ts{\textstyle}
\def\cl{\centerline}
\def\eq{\eqno}

\def\BZT{{\rm Z{\hbox to 3pt{\hss\rm Z}}}}
\def\BZS{{\hbox{\sevenrm Z{\hbox to 2.3pt{\hss\sevenrm Z}}}}}
\def\BZSS{{\hbox{\fiverm Z{\hbox to 1.8pt{\hss\fiverm Z}}}}}
\def\BZ{{\mathchoice{\BZT}{\BZT}{\BZS}{\BZSS}}}
\def\BQT{\,\hbox{\hbox to -2.8pt{\vrule height 6.5pt width .2pt\hss}\rm Q}}
\def\BQS{\,\hbox{\hbox to -2.1pt{\vrule height 4.5pt width .2pt\hss}$
    \scriptstyle\rm Q$}}
\def\BQSS{\,\hbox{\hbox to -1.8pt{\vrule height 3pt width
    .2pt\hss}$\scriptscriptstyle \rm Q$}}

\def\BCT{\,\hbox{\hbox to -3pt{\vrule height 6.5pt width .2pt\hss}\rm C}}
\def\BCS{\,\hbox{\hbox to -2.2pt{\vrule height 4.5pt width .2pt\hss}$
    \scriptstyle\rm C$}}
\def\BCSS{\,\hbox{\hbox to -2pt{\vrule height 3.3pt width
    .2pt\hss}$\scriptscriptstyle \rm C$}}

{\nopagenumbers
\rm
\baselineskip=0.6cm
\smno
\cl{{BONN-HE-92-23 August 1992}}
\vskip 1truecm
\cl{\bf{\gro N=2 supersymmetric W-algebras }}
\vskip 1.3truecm
\cl{{\mit{Ralph Blumenhagen}}
\footnote{$^*$}{e-mail: unp039@ibm.rhrz.uni-bonn.de} }
\vskip 1truecm
\cl{\sl Physikalisches Institut der Universit\"at Bonn}
\cl{\sl Nu\ss allee 12, D-5300 Bonn 1}
\cl{\sl Germany}
\vskip 2.5truecm
\cl{\bf Abstract }
\bigno
{\leftskip=1 truecm\rightskip=1 truecm
We investigate extensions of the N=2 super Virasoro algebra by
one additional super primary field and its charge conjugate.
Using a supersymmetric covariant
formalism we construct all N=2 super W-algebras up to spin 5/2
of the additional generator. Led by these first examples
we close with some conjectures on the
classification of N=2 ${\cal SW}(1,\Dt)$ algebras.\smno  }
\vfill\eject }
\pageno=1
\rm
\baselineskip=0.6cm
\smno
{\cl{\bf 1. Introduction}}
\meno
In 1987 Gepner [1-3] proposed a huge class of new N=2 supersymmetric string
vacua by assuming that compactification of the heterotic string
to four dimensions leads to a minimal N=2 theory with total central
charge $c=9$. It was shown in ref. [4] that N=2 world-sheet supersymmetry
is necessary for N=1 space-time supersymmetry. For the
internal $c=9$ theory Gepner takes tensor products of minimal, unitary
representations of the N=2 super Virasoro algebra [5]. Surprisingly,
these algebraically
defined  theories correspond geometrically to compactifications
on Calabi-Yau spaces [2,3].\pano
The GUT gauge groups of these models are $E_8\times E_6\times U(1)^{r-1}$
where {\it r} denotes the number of tensored N=2 models. Since these
additional $U(1)$ groups are not observed in nature one wants to get rid
of them. This motivated Kazama-Suzuki [6,7] to search for N=2 theories
having minimal representations for $c=9$. They constructed and classified
coset models
of N=1 super Kac-Moody algebras which have an extended N=2 supersymmetry.
Several models with the desired value $c=9$ were found.
 The symmetry algebras of these
Kazama-Suzuki models should be generically existing N=2 super W-algebras
[8]. The above application is the most important reason to classify
N=2 supersymmetric W-algebras.
\pano We denote the extension of the N=2 super Virasoro algebra
by super primary fields of dimension $\Dt_1\ldots\Dt_n$ by
${\cal SW}(1,\Dt_1\ldots\Dt_n)$. Note that usually the primaries consist of
pairs
of opposite $U(1)$ charge $Q$. The motivation for our work was to
find new extensions of the N=2 super Virasoro algebra by explicit
construction.
{}From the recent extensive work on N=0 and N=1 (super) W-algebras [9-12]
one knows
of the occurence of (super) W-algebras which exist only for a few values
of the central charge $c$. The hope is that nature may have chosen
such an isolated N=2 super W-symmetry for the vacuum.
\pano
Up to now only two examples of nonlinear N=2 super W-algebras
are known explicitly, ${\cal SW}(1,3/2)$ [13] and ${\cal SW}(1,2)$ [14].
For the first
one Inami et al., using a conformal bootstrap algorithm, found
that all four point functions are associative for generic value of $c$.
But they claimed that carrying out an algebraic calculation and checking
Jacobi identities would yield stronger restrictions on $c$.
Our calculations confirm this conjecture. The special case $c=9$ and
$Q=3$, in which the additional, chiral spin $3/2$ generator is interpreted
as the spectral flow operator, has been studied by Odake [15,16] intensively.
The second algebra has only been
constructed for vanishing $U(1)$ charge of the spin two generator.
Romans found that this algebra exists for generic value of $c$.\pano
In this paper we continue the series of ${\cal SW}(1,\Dt)$ algebras
up to spin three. The first four examples disclose already enough structure
to speculate on the classification of these algebras.
We conjecture the existence of a new N=2 super W-algebra existing
for $c=9$, the ${\cal SW}(1,11/2)$ algebra. \pano
The paper is organized as follows: In the second section we develop
an N=2 supersymmetric covariant formalism for the construction
of N=2 super W-algebras. We show that requiring invariance under
super M\"obius transformations puts severe restrictions on the
structure of these algebras. We also define supersymmetric
normal ordered products. It is this structural investigation that
makes the explicit construction of super W-algebras feasible.\pano
In the third section we apply the whole formalism to the construction
of ${\cal SW}(1,\Dt)$ algebras. Using the
symbolic manipulation package REDUCE, we extend the series of known
W-algebras up to $\Dt=5/2$. For $\Dt=3$ we are able to give
necessary conditions for existence. Different from the N=0 and  N=1
case, ${\cal SW}(1,\Dt)$ algebras are also allowed to exist for $c\ge3$.
\pano
In the fourth section we discuss the results obtained and present
some speculations about the classification of
${\cal SW}(1,\Dt)$ algebras. We conjecture for example
all ${\cal SW}(1,\Dt)$ algebras to exist for $c=3$ and $Q=\pm 1$ if
$\Dt$ is half-integer.\pano
The fifth section concludes this paper presenting a short summary.
\meno
\bigno
{\cl{\bf 2. N=2 supersymmetric CFT}}
\bigno
In this section we generalize the formulae which have been presented
for the case N=0 and N=1 in ref. [9,15]. We will see
that the structure of N=2 W-algebras is considerably restricted by
invariance
under N=2 super M\"obius transformations. Since the aim of our work
is the covariant construction of N=2 super W-algebras, we have to
express all involved fields as so-called N=2 super quasiprimary fields.
Thus a supersymmetric notion of normal ordering has to be worked out.
\vfill\eject\pano
{2.1. SU(1,1) INVARIANT CFT} \meno
First we review the basic structural property of SU(1,1) invariant chiral
algebras. For a more detailed exposition we refer the reader to ref. [9]
and ref. [15] for the N=1 supersymmetric case.\meno
We define a special class of chiral fields
\meno
$$\varphi(z)=\sum_{n\in \BZ} \varphi_n z^{n-d} \eqno(2.1)$$
\meno
with respect to the energy-momentum tensor
\meno
$$L(z)=\sum_{n\in \BZ} L_n z^{n-2} \eqno(2.2) $$
\meno
by requiring that
\meno
$$\lbrack L_m,\varphi_n\rbrack =(n-(d-1)m)\varphi_{m+n}
\eqno(2.3)$$ \meno
is satisfied for $m\in \lbrace -1,0,1 \rbrace$. These fields are called
{\it quasiprimary}.\pano
The Fourier-modes of a set  $\lbrace\varphi_i(z)\rbrace$
of such fields have a simple Lie algebra
structure
\meno
$$\lbrack\varphi_{i,m},\varphi_{j,n}\rbrack=\sum_k
           C_{ij}^{k}\>p_{ijk}(m,n)\>\varphi_{k,m+n}+d_{ij}\delta_{n,-m}
           {n+h(i)-1\choose{2h(i)-1}}, \eqno(2.4) $$
\meno
where the structure constants $d_{ij}$ and $C_{ijk}$ are given by
\meno
$$d_{ij}=\la 0|\varphi_{i,-h(i)}\varphi_{j,h(j)}|0\ra , \eqno(2.5)$$
\smno
$$C_{ijk}=\la 0|\varphi_{k,-h(k)}\varphi_{i,h(k)-h(j)}\varphi_{j,h(j)}|0\ra
               \eqno(2.6)$$
and
\meno
$$C_{ij}^l d_{lk} = C_{ijk}.\eq(2.7)$$ \bigno
\meno
The polynomials $p_{ijk}(m,n)$ depend
only on the conformal dimensions of the involved fields and are explicitly
known [9].\pano
Furthermore, we introduced quasiprimary normal ordered products
${\cal N}(\varphi_j\partial^n\varphi_i)$ by a qua\-si\-primary projection of
the natural normal ordered products
$N(\varphi_j\partial^n\varphi_i)$ occuring in the regular part of the
operator product expansion (OPE).
With these structural concepts
we were able to construct a whole bunch of new W-algebras.
\bigno
{2.2. N=2 SUPERCONFORMAL ALGEBRA} \meno
In this subsection we review some elementary notions of N=2
supergeometry [16,17],
and we will define a special class of superfields in N=2 superspace.
Because we are dealing with left- or right-handed field theories we
treat in the following only one chirality.
\meno
The N=2 superspace is the extension of the complex plane by two
Grassmann variables. A point is a triple $Z=(z,\theta,\o{\theta})$
where $\theta,\o{\theta}$ are anticommuting objects.\pano
On this superspace one defines two covariant derivatives
\meno
$$D = \partial_{\theta} +{1\over 2}\o{\theta}\partial_z,  \eqno(2.8) $$
\meno
$$\o{D} = \partial_{\o{\theta}} +{1\over 2}{\theta}\partial_z \eqno(2.9) $$
\meno
which are nilpotent and satisfy
$$ \lbrace D,\o{D}\rbrace =\partial_z. \eqno(2.10)$$
\meno
Furthermore, we introduce the super translationally invariant
super interval
$Z_{ij}$ of two points $Z_i$ and $Z_j$
\meno
$$Z_{ij}= z_i-z_j-{1\over 2}(\theta_i\o\theta_j+\o\theta_i\theta_j).
\eqno(2.11)$$
\meno
One obtains three other translationally invariant intervals by multiplying
$Z_{ij}$ by $\theta_{ij}=\theta_i-\theta_j$,
$\o{\theta}_{ij}=\o{\theta}_i-\o{\theta}_j$ and
$\theta_{ij}\o{\theta}_{ij}$.
\pano
Similar to the complex plane, one can define the N=2 super conformal group
acting on N=2 superspace. The generators of this group form an
infinite dimensional Lie algebra which admits a central extension.
The resulting algebra is the so-called N=2 super Virasoro algebra
\bigno
$$\eqalignno{\lb L_m,L_n \rb&=(n-m)L_{m+n}+{c\over 12}(n^3
            -n)\delta_{m,-n} &\cr
   \lb L_m,G_r \rb&=(r-{\ts{m\over 2}})G_{m+r} &\cr
\lb L_m,\o{G}_r \rb&=(r-{\ts{m\over 2}})\o{G}_{m+r} &\cr
\lb L_m,J_n \rb&=nJ_{m+n} &\cr
\lb J_m,J_n \rb&={c\over 3}n\delta_{m,-n} &(2.12)\cr
\lb J_m,G_r \rb&=G_{m+r} &\cr
\lb J_m,\o{G}_r \rb&=-\o{G}_{m+r} &\cr
\lbrace G_r,\o{G}_s \rbrace&=2L_{r+s}+(s-r)J_{r+s}
           +{2c\over 3}{s+{1\over 2}\choose 2} \delta_{r,-s}&\cr
\lbrace G_r,G_s \rbrace&=\lbrace \o{G}_r,\o{G}_s \rbrace=0. &\cr}$$
\bigno
The essential property of this algebra is the appearance of the
abelian current $J(z)$. The fields
\bigno
$$\eqalignno{ L(z)&=\sum_{n\in Z}L_n z^{n-2}\quad
              \quad J(z)=\sum_{n\in Z}J_n z^{n-1} &(2.13)\cr
   G(z)&=\sum_r G_r z^{r-{3\over 2}}\quad\quad
   \o{G}(z)=\sum_r \o{G}_r z^{r-{3\over 2}} &\cr }$$
\meno
can be put together into one N=2 superfield
\meno
$$ {\cal L}(Z) = J(z)+\ts{1\over\sqrt 2}(\theta \o{G}(z)
    -\o{\theta}G(z))+\theta\o{\theta}L(z).\eqno(2.14) $$
\meno
The N=2 super Virasoro algebra has been studied very intensively during
the last years, and the unitary minimal representations of this algebra
are well known [5]. These exist for a discrete series of
$c$ values
\meno
$$ c={3k\over k+2}, \quad\quad\quad k\in \BZ_+.\eqno(2.15)$$
\meno
We will review the minimal representations of this algebra in the fourth
section  where we discuss the results of our work.
{}From now on we will only consider the Neveu-Schwarz sector, i.e.
we choose periodic boundary conditions for all fermionic fields
($r\in \BZ+1/2$ in (2.13)). \pano
The superconformal algebra contains the subalgebra of super M\"obius
transformations which are generated by
$\lbrace L_1, L_0, L_{-1}, G_{1/2}, G_{-1/2},
 \o{G}_{1/2}, \o{G}_{-1/2}, J_0\rbrace$.
This algebra is isomorphic to the super Lie algebra $Osp(2|2)$ and
is analogous to $SU(1,1)$ in the N=0 case or to $Osp(1|2)$ in the N=1 case.
In the rest of this section, we will show that requiring an $Osp(2|2)$
invariant superconformal field theory puts strong restrictions on its
structure. \pano
The maximal abelian subalgebra of the superconformal algebra is
generated by $L_0$ and $J_0$. Thus each state in the Hilbert space of a
superconformal field theory has at least two quantum numbers, the
conformal weight $\Dt$ and the $U(1)$ charge $Q$. Because of the
isomorphism between states and fields, the definition of super primary
fields involves also two parameters $(\Dt,Q)$. If a superfield
$\Phi_{\Dt}^{Q}(Z)$
satisfies the following equations then it is called a {\it super primary
field} of dimension $\Dt$ and charge $Q$:\bigno
$$\eqalignno{ \lbrack L_m,\Phi(Z)\rbrack &=z^{-m}\Bigl[z\partial_z-
     {\ts{(m-1)\over 2}}\theta\partial_{\theta}-
     {\ts{(m-1)\over 2}}\o{\theta}\partial_{\o{\theta}} &\cr
      &\quad\quad\quad -\Dt(m-1)+
     {\ts {Q\over 4}}m(m-1)\theta\o{\theta}z^{-1}\Bigr]\Phi(Z) &\crns
     \lbrack G_r,\Phi(Z)\rbrack_{\pm} &=\sqrt 2 z^{-r+1/2}\Bigl[
     \partial_{\o\theta}-{\ts {1\over 2}}\theta\partial_z-
    (\Dt+{\ts {Q\over 2}})(r- {\ts {1\over 2}})\theta z^{-1} &(2.16)\cr
     &\quad\quad\quad -{\ts {1\over 2}}(r- {\ts {1\over 2}})\theta
\o{\theta}z^{-1}
     \partial_{\o\theta}\Bigr]\Phi(Z) &\crns
     \lbrack \o{G}_r,\Phi(Z)\rbrack_{\pm} &=\sqrt 2 z^{-r+1/2}\Bigl[
     \partial_{\theta}-{\ts {1\over 2}}\o{\theta}\partial_z-
    (\Dt-{\ts {Q\over 2}})(r- {\ts {1\over 2}})\o{\theta}z^{-1} &\cr
     &\quad\quad\quad +{\ts {1\over 2}}(r- {\ts {1\over 2}})\theta\o{\theta}
z^{-1}
     \partial_{\theta}\Bigr]\Phi(Z) &\crns
    \lbrack J_m,\Phi(Z)\rbrack &=z^{-m}\bigl[-\theta\partial_{\theta}
     +\o\theta\partial_{\o\theta}-\Dt m\theta\o\theta z^{-1}+
      Q\bigr]\Phi(Z) &\cr }$$
\meno
where for two fermionic fields one takes the anticommutator, the
commutator otherwise.\pano
If (2.16) is satisfied only with respect to the subalgebra $Osp(2|2)$,
$\Phi(Z)$ is called a {\it super quasiprimary field}. The super
Virasoro field defined above is an example for a super quasiprimary
field of dimension one and charge zero.\pano
Note that $G_{1/2}$ and $\o{G}_{1/2}$ are the generators of
supertranslations on
the space of fields\meno
$$\lbrack G_{1/2},{\Phi}(Z)\rbrack_{\pm} =\sqrt 2 (\partial_{\o\theta}
    -{\ts{1\over 2}}\theta\partial_z)\Phi(Z)\eqno(2.17) $$
\meno
$$\lbrack \oG_{1/2},{\Phi}(Z)\rbrack_{\pm} =\sqrt 2 (\partial_{\theta}
    -{\ts{1\over 2}}\o\theta\partial_z)\Phi(Z) \eqno(2.18)$$
\meno
whereas $J_0$ generates a "deformed" rotation in the pure Grassmann-plane
\meno
$$\lbrack J_{0},{\Phi}(Z)\rbrack_{\pm} = (-\theta\partial_{\theta}
    +\o\theta\partial_{\o\theta}+Q)\Phi(Z).\eqno(2.19) $$
\meno
These transformations leave invariant
the four super intervals defined in (2.11), if the charges $Q=1,-1$ are
assigned to  $\theta$ and $\o\theta$, respectively.
By expanding the superfields in a Taylor series with respect to $\theta$
and $\o\theta$ one obtains four components
\meno
$$\Phi(Z) = \varphi(z)+{\ts{1\over\sqrt 2}}(\theta\o\psi(z)
             -\o\theta\psi(z))+\theta\o\theta\chi(z) \eqno(2.20)    $$
\pano or \pano
$$\Phi(Z) = \Phi_1(z)+{\ts{1\over\sqrt 2}}(\theta\Phi_{\o\theta}(z)
             -\o\theta\Phi_{\theta}(z))
              +\theta\o\theta\Phi_{\theta\o\theta}(z) \eqno(2.20)    $$
\meno
and realizes that the "highest" component $\chi(z)$ is not quasiprimary
with respect to the Virasoro algebra if the charge $Q$ is different from
zero. Thus, for component calculations we will use instead its
quasiprimary projection
\meno
$$\chi'(z)=\chi(z)-{\ts{Q\over{4\Dt}}}\partial_z\varphi(z).\eqno(2.21)$$
\meno
Hence a super (quasi)primary field is defined by the following
commutation relations
\bigno
$$\eqalignno{ \lbrack L_m,\varphi_n\rbrack_{\pm} &=(n-(\Dt-1)m)
           \varphi_{m+n} &\cr
           \lbrack L_m,\psi_n\rbrack_{\pm} &=(n-(\Dt-{\ts{1\over 2}})m)
           \psi_{m+n} &\cr
           \lbrack L_m,\o\psi_n\rbrack_{\pm} &=(n-(\Dt-{\ts{1\over 2}})m)
           \o\psi_{m+n} &\cr
           \lbrack L_m,\chi'_n\rbrack_{\pm} &=(n-\Dt m)
           \chi'_{m+n} &\cr
          \lbrack G_m,\varphi_n\rbrack_{\pm} &=-\psi_{m+n} &\cr
          \lbrack \oG_m,\varphi_n\rbrack_{\pm} &=\o\psi_{m+n} &\cr
          \lbrack G_m,\psi_n\rbrack_{\pm} &=
           \lbrack \oG_m,\o\psi_n\rbrack_{\pm} =0 &\cr
          \lbrack G_m,\o\psi_n\rbrack_{\pm} &=(1+{\ts{Q\over2\Dt}})
             (n-(2\Dt+1)m) \varphi_{m+n}  +2\chi'_{m+n} &(2.22)\cr
          \lbrack \oG_m,\psi_n\rbrack_{\pm} &=-(1-{\ts{Q\over2\Dt}})
             (n-(2\Dt+1)m) \varphi_{m+n}  +2\chi'_{m+n} &\cr
          \lbrack G_m,\chi'_n\rbrack_{\pm} &={\ts{1\over 2}}
           (1+{\ts{Q\over{2\Dt}}})(n-2\Dt m)\psi_{m+n} &\cr
          \lbrack \oG_m,\chi'_n\rbrack_{\pm} &={\ts{1\over 2}}
           (1-{\ts{Q\over{2\Dt}}})(n-2\Dt m)\psi_{m+n} &\cr
          \lbrack J_m,\varphi_n\rbrack_{\pm} &=Q\varphi_{m+n} &\cr
          \lbrack J_m,\psi_n\rbrack_{\pm} &=(Q+1)\psi_{m+n} &\cr
          \lbrack J_m,\o\psi_n\rbrack_{\pm} &=(Q-1)\o\psi_{m+n} &\cr
          \lbrack J_m,\chi_n\rbrack_{\pm} &=-\Dt(1+{\ts{Q^2\over 4\Dt^2}})
               m\varphi_{m+n}+Q\chi'_{m+n}. &\cr   }$$
\bigno
These equations imply that one can obtain the Fourier-modes of the
higher components
of a super quasiprimary field if one knows that of its lowest component
$\varphi_{\Dt}|0\ra$
\meno
$$\eqalignno{ \psi_{\Dt+1/2}|0\ra &=-G_{1/2}\varphi_{\Dt}|0\ra\quad\quad
           \o\psi_{\Dt+1/2}|0\ra =\o{G}_{1/2}\varphi_{\Dt}|0\ra &(2.23)\cr
           \chi'_{\Dt+1}|0\ra &={\ts{ {1\over 2}(G_{1/2}\o{G}_{1/2}
            -(1+{Q\over 2\Dt})L_1) }} \varphi_{\Dt}|0\ra. &\cr }$$
\meno
During the explicit construction of  W-algebras we will use these relations
to determine the higher components of super quasiprimary normal ordered
products.\pano
Now we are in a position to analyse
n-point functions and the OPE structure for super quasiprimary fields.
We will show that the OPE of two such superfields is completely
determined up to some coupling constants. \bigno\meno
{2.3 TWO- AND THREE-POINT FUNCTIONS}\meno
In this subsection we review and improve the expressions known in the
literature, which are incomplete or apply only to some specific cases.
(Nevertheless, see ref. [17,18] for a more detailed derivation.)
\pano
We assume that in the theory there is a vacuum invariant under
super M\"obius transformations. By taking
commutators with the eight generators of $Osp(2|2)$ inside the
vacuum correlators we obtain eight
super differential equations for the n-point functions, which, however,
are not all independent because of (2.12).
For example, the two-point function obeys the following differential equation
which results from taking the commutator with $J_0$:\meno
$$  \sum_{i=1}^2  ( -\theta_i\partial_{\theta_i}
    +\o\theta_i\partial_{\o\theta_i}+Q_i)\la\Phi (Z_1)\Phi(Z_2)\ra=0.
\eqno(2.24) $$
\meno
The solution of these differential equations for the two-point function is
\meno
$$ \la\Phi_{\Dt_i}^{Q_i} (Z_1)\Phi_{\Dt_j}^{Q_j}(Z_2)\ra=
    {D_{ij}\over Z_{12}^{2\Dt_i}}\left(1-{Q_i\over 2}{\theta_{12}
\o\theta_{12}
     \over  Z_{12}} \right) \delta_{\Dt_i,\Dt_j}
      \delta_{Q_i,-Q_j}. \eqno(2.25) $$
\meno
Thus, only the two-point function for fields of opposite charge does not
vanish. \pano
The three-point functions are more complicated since three cases have
to be distinguished.\smno\vfill\eject\pano
\hskip 1cm (i) If $Q_i+Q_j=Q_k$  then
\meno
$$\eqalignno{ \la \Phi_{\Dt_k}^{-Q_k}(Z_1)\>\Phi_{\Dt_i}^{Q_i}(Z_2)\>
\Phi_{\Dt_j}^{Q_j}(Z_3)\ra &=
C_{ijk}2\left[
     {\ts{ (1-{Q_i\over 2}{\theta_{23}\o\theta_{23} \over  Z_{23}}) }}
     {\ts{ (1+{Q_k\over 2}{\theta_{13}\o\theta_{13} \over  Z_{13}}) }}
{\ts{ {1\over Z_{12}^{\gamma_1}\> Z_{23}^{\gamma_2} \> Z_{13}^{\gamma_3}}
    }} \right] &\cr
&+C_{ijk}\alpha_{ijk}\gamma_1 \Bigl[
     {\ts{ ({\theta_{23} \over  Z_{23}}-{\theta_{13}\over Z_{13}}) }}
     {\ts{ ({\o\theta_{23} \over  Z_{23}}-{\o\theta_{13}\over Z_{13}}) }}
{\ts{ {1\over Z_{12}^{\gamma_1+1}\> Z_{23}^{\gamma_2-1} \>
Z_{13}^{\gamma_3-1}}
    }}  &\cr
&\quad\quad\quad\quad\quad -{\ts{ \theta_{23}\o\theta_{23}\theta_{13}
\o\theta_{13}\over 2}}
   {\ts{ ({Q_i \over  Z_{13}}-{Q_k\over Z_{13}}) }}
{\ts{ {1\over Z_{12}^{\gamma_1+1}\> Z_{23}^{\gamma_2} \> Z_{13}^{\gamma_3}}
   }} \Bigr]  &(2.26)\cr   }$$
\bigno
where $\gamma_1=\Dt_k+\Dt_i-\Dt_j$,  $\gamma_2=\Dt_i+\Dt_j-\Dt_k$
and $\gamma_3=\Dt_k+\Dt_j-\Dt_i$.\pano
$C_{ijk}$ and $\alpha_{ijk}$ are independent parameters in contrast
to ref. [17,18].
\meno
\hskip 1cm (ii) If $Q_i+Q_j=Q_k-1$  then
\meno
$$\eqalignno{ \la \Phi_{\Dt_k}^{-Q_k}(Z_1)\>\Phi_{\Dt_i}^{Q_i}(Z_2)\>
\Phi_{\Dt_j}^{Q_j}(Z_3)\ra &=
{\ts{ {\ds{ C_{ijk} }}\over
 Z_{12}^{\gamma_1+{1\over 2}}\> Z_{23}^{\gamma_2-{1\over 2}}
   \> Z_{13}^{\gamma_3-{1\over 2}}} }\left[
  {\ts{ ({\o\theta_{23} \over  Z_{23}}-{\o\theta_{13}\over Z_{13}}) }}
   +{\ts{ (Q_i\o\theta_{13}\theta_{23}\o\theta_{23}+Q_k\o\theta_{23}
     \theta_{13}\o\theta_{13})\over 2 Z_{23}\> Z_{13} }}\right] &\cr
    & &(2.27)\cr }$$
\bigno
\hskip 1cm (iii) If $Q_i+Q_j=Q_k+1$  then
\meno
$$\eqalignno{ \la \Phi_{\Dt_k}^{-Q_k}(Z_1)\>\Phi_{\Dt_i}^{Q_i}(Z_2)\>
\Phi_{\Dt_j}^{Q_j}(Z_3)\ra &=
{\ts{ {\ds{ C_{ijk} }}\over
 Z_{12}^{\gamma_1+{1\over 2}}\> Z_{23}^{\gamma_2-{1\over 2}}
   \> Z_{13}^{\gamma_3-{1\over 2}}} }\left[
  {\ts{ ({\theta_{23} \over  Z_{23}}-{\theta_{13}\over Z_{13}}) }}
   +{\ts{ (Q_i\theta_{13}\theta_{23}\o\theta_{23}+Q_k\theta_{23}
     \theta_{13}\o\theta_{13})\over 2 Z_{23}\> Z_{13} }}\right] &\cr
    & &(2.28)\cr }$$
\meno
In (ii) and (iii) there exists only one independent coupling constant
$C_{ijk}$. These rather lengthy three-point functions contain all
the information we need for deriving N=2 supersymmetric
OPEs.
\bigno
2.4. N=2 SUPERSYMMETRIC OPE  \meno
In this subsection we derive the general form
of the OPE of two $Osp(2|2)$ invariant superfields,
using the results of the last subsection.
To this end, we write down the most general ansatz for an OPE which is
invariant
under the operation of the generators ${L_0,L_1,G_{1/2},\oG_{1/2},J_0}$.
\meno
\hskip 1cm (i) If $Q_i+Q_j=Q_k$  then
$$\eqalignno{  \Phi_{\Dt_i}^{Q_i}(Z_2)\>\Phi_{\Dt_j}^{Q_j}(Z_3) &=
   \sum_{n=0}^{\infty} {\ts{ C_{ij}^k A_{ijk}^n\over Z_{23}^{\Dt-n} }}
   \partial_z^n \Phi_{\Dt_k}^{Q_k}(Z_3)+
   \sum_{n=1}^{\infty} {\ts{ C_{ij}^k B_{ijk}^n\over Z_{23}^{\Dt-n} }}
   [\o{D},D]\partial_z^{n-1} \Phi_{\Dt_k}^{Q_k}(Z_3) &\cr
 &+  \sum_{n=0}^{\infty} {\ts{ C_{ij}^k C_{ijk}^n\theta_{23}\over
Z_{23}^{\Dt-n} }}
   D\partial_z^{n} \Phi_{\Dt_k}^{Q_k}(Z_3) +
   \sum_{n=0}^{\infty} {\ts{ C_{ij}^k D_{ijk}^n\o\theta_{23}\over
Z_{23}^{\Dt-n} }}
  \o{D}\partial_z^{n} \Phi_{\Dt_k}^{Q_k}(Z_3) &(2.29)\cr
 &+  \sum_{n=-1}^{\infty} {\ts{ C_{ij}^k E_{ijk}^n\theta_{23}\o\theta_{23}
\over Z_{23}^{\Dt-n} }}
   \partial_z^{n+1} \Phi_{\Dt_k}^{Q_k}(Z_3) +
   \sum_{n=0}^{\infty} {\ts{ C_{ij}^k F_{ijk}^n\theta_{23}\o\theta_{23}
\over Z_{23}^{\Dt-n} }}
   [\o{D},D]\partial_z^{n} \Phi_{\Dt_k}^{Q_k}(Z_3) &\cr }$$
\meno
The coefficients in this expansion can be determined in the following
way. One inserts the OPE into the three point function and
using the two point function (2.25), obtains
a rational function in the variables
$Z_{23}$, $Z_{13}$, $\theta_{23}$, $\theta_{13}$, $\o\theta_{23}$ and
$\o\theta_{13}$. This expression must be equal to the general three point
function (2.26) after having fixed the points in a special way by means of
an appropriate super M\"obius transformation.
For example, one may shift the three
points to $z_1=\infty$, $z_2=1$, $z_3=0$, $\theta_{23}=0$ and
$\o\theta_{23}=0$. Comparing the two expressions yields a linear
equation for the coefficients in the OPE (2.29).
After a rather lengthy calculation, one can generate six independent
equations which can be solved easily in terms of the free parameter
$\alpha=\alpha_{ijk}$, the conformal dimensions and the $U(1)$ charges:
\meno
$$\eqalignno{ A_{ijk}^n&={2\over n!} {2\Dt_k+n\choose n}^{-1}
           {\gamma_1+n-1\choose n}\left(1+{n\over 2}{Q_k/2-\alpha\Dt_k\over
           (Q_k/2)^2-\Dt_k^2}\right) &\crnm
           B_{ijk}^n&={1\over (n-1)!} {2\Dt_k+n\choose n}^{-1}
           {\gamma_1+n-1\choose n}\left({Q_k/2-\alpha\Dt_k\over
           (Q_k/2)^2-\Dt_k^2}\right) &\crnm
           C_{ijk}^n&={2\Dt_k\over n!} {2\Dt_k+n\choose n+1}^{-1}
           {\gamma_1+n\choose n+1}\left({1+\alpha\over
           Q_k/2+\Dt_k}\right) &\cr &&(2.30)\cr
           D_{ijk}^n&=-{2\Dt_k\over n!} {2\Dt_k+n\choose n+1}^{-1}
           {\gamma_1+n\choose n+1}\left({1-\alpha\over
           Q_k/2-\Dt_k}\right) &\crnm
           E_{ijk}^n&={1\over (n+1)!} {2\Dt_k+n+1\choose n+1}^{-1}
         \biggl\{ {\gamma_1+n+1\choose n+2}{(n+2)\over 2}\left[2\alpha+
          (n+1)\left({Q_k/2-\alpha\Dt_k\over
           (Q_k/2)^2-\Dt_k^2}\right)\right]  &\cr
        & \quad\quad\quad\quad -{ Q_i}{\gamma_1+n\choose n+1}
          \left[ 1+{(n+1)\over 2} \left({\alpha Q_k/2-\Dt_k\over
           (Q_k/2)^2-\Dt_k^2}\right)\right] \biggr\} &\crnm
           F_{ijk}^n&=-{Q_i\over 2n!} {2\Dt_k+n+1\choose n+1}^{-1}
          {\gamma_1+n\choose n+1}\left({Q_k/2-\alpha\Dt_k\over
           (Q_k/2)^2-\Dt_k^2}\right)  &\cr
          &\quad +{\Dt_k\over n!} {2\Dt(k)+n+1\choose n+2}^{-1}
          {\gamma_1+n+1\choose n+2}\left({\alpha Q_k/2-\Dt_k\over
           (Q_k/2)^2-\Dt_k^2}\right)  &\cr  }$$
\meno
The coupling constant $C_{ij}^k$ is then determined by the linear
system $C_{ij}^l D_{lk}=C_{ijk}$.
\pano
Before presenting some simple examples, we give the OPEs for the two other
cases.
\meno
\hskip 1cm (ii) If $Q_i+Q_j=Q_k-1$  then
$$\eqalignno{  \Phi_{\Dt_i}^{Q_i}(Z_2)\>\Phi_{\Dt_j}^{Q_j}(Z_3) &=
   \sum_{n=0}^{\infty} {\ts{ C_{ij}^k A_{ijk}^n\o\theta_{23}
   \over Z_{23}^{\Dt-n+1/2} }}
   \partial_z^n \Phi_{\Dt_k}^{Q_k}(Z_3)+
   \sum_{n=1}^{\infty} {\ts{ C_{ij}^k B_{ijk}^n\over Z_{23}^{\Dt-n+1/2} }}
   D\partial_z^{n-1} \Phi_{\Dt_k}^{Q_k}(Z_3) &(2.31)\cr
  &+ \sum_{n=1}^{\infty} {\ts{ C_{ij}^k C_{ijk}^n\o\theta_{23}
   \over Z_{23}^{\Dt-n+1/2} }}  [\o{D},D]
   \partial_z^{n-1} \Phi_{\Dt_k}^{Q_k}(Z_3)+
   \sum_{n=0}^{\infty} {\ts{ C_{ij}^k D_{ijk}^n\theta_{23}\o\theta_{23}
   \over Z_{23}^{\Dt-n+1/2} }}
   D\partial_z^{n} \Phi_{\Dt_k}^{Q_k}(Z_3) &\cr   }$$
\bigno
$$\eqalignno{ A_{ijk}^n&={1\over n!} {2\Dt_k+n\choose n}^{-1}
           {\gamma_1+n-1/2\choose n}\left(1+{n\over 2(\Dt_k+Q_k/2)}
           \right) &\crnm
          B_{ijk}^n&={1\over (n-1)!} {2\Dt_k+n-1\choose n-1}^{-1}
           {\gamma_1+n-3/2\choose n-1}{1\over (\Dt_k+Q_k/2)}&\cr &&(2.32)\cr
          C_{ijk}^n&={1\over (n-1)!} {2\Dt_k+n\choose n}^{-1}
           {\gamma_1+n-1/2\choose n}{1\over 2(\Dt_k+Q_k/2)} &\crnm
          D_{ijk}^n&=-{1\over n!} {2\Dt_k+n\choose n}^{-1}
            {1\over (\Dt_k+Q_k/2)}\left[{Q_i\over 2}
           {\gamma_1+n-1/2\choose n}+{(n+1)\over 2}
           {\gamma_1+n+1/2\choose n+1} \right] &\cr }$$
\bigno
\hskip 1cm (iii) If $Q_i+Q_j=Q_k+1$  then
$$\eqalignno{  \Phi_{\Dt_i}^{Q_i}(Z_2)\>\Phi_{\Dt_j}^{Q_j}(Z_3) &=
   \sum_{n=0}^{\infty} {\ts{ C_{ij}^k A_{ijk}^n\theta_{23}
   \over Z_{23}^{\Dt-n+1/2} }}
   \partial_z^n \Phi_{\Dt_k}^{Q_k}(Z_3)+
   \sum_{n=1}^{\infty} {\ts{ C_{ij}^k B_{ijk}^n\over Z_{23}^{\Dt-n+1/2} }}
   \o{D}\partial_z^{n-1} \Phi_{\Dt_k}^{Q_k}(Z_3) &(2.33)\cr
  &+ \sum_{n=1}^{\infty} {\ts{ C_{ij}^k C_{ijk}^n\theta_{23}
   \over Z_{23}^{\Dt-n+1/2} }}  [\o{D},D]
   \partial_z^{n-1} \Phi_{\Dt_k}^{Q_k}(Z_3)+
   \sum_{n=0}^{\infty} {\ts{ C_{ij}^k D_{ijk}^n\theta_{23}\o\theta_{23}
   \over Z_{23}^{\Dt-n+1/2} }}
   \o{D}\partial_z^{n} \Phi_{\Dt_k}^{Q_k}(Z_3) &\cr   }$$
\bigno
$$\eqalignno{ A_{ijk}^n&={1\over n!} {2\Dt_k+n\choose n}^{-1}
           {\gamma_1+n-1/2\choose n}\left(1+{n\over 2(\Dt_k-Q_k/2)}
           \right) &\crnm
          B_{ijk}^n&={1\over (n-1)!} {2\Dt_k+n-1\choose n-1}^{-1}
           {\gamma_1+n-3/2\choose n-1}{1\over (\Dt_k-Q_k/2)}&\cr &&(2.34)\cr
          C_{ijk}^n&=-{1\over (n-1)!} {2\Dt_k+n\choose n}^{-1}
           {\gamma_1+n-1/2\choose n}{1\over 2(\Dt_k-Q_k/2)} &\crnm
          D_{ijk}^n&=-{1\over n!} {2\Dt_k+n\choose n}^{-1}
            {1\over (\Dt_k-Q_k/2)}\left[{Q_i\over 2}
           {\gamma_1+n-1/2\choose n}-{(n+1)\over 2}
           {\gamma_1+n+1/2\choose n+1} \right]. &\cr }$$
\bigno
After the technical considerations of this subsection, we are now in
the position to express the main relations of the super Virasoro
algebra in a very compact form. First,
application of the formula (2.30) to the OPE of two super Virasoro fields
yields
\meno
$$\eqalignno{ {\cal L}(Z_2)\>{\cal L}(Z_3) &={c/3 \over Z_{23}^2}+
     {\theta_{23}\o\theta_{23}\over Z_{23}^2}{\cal L}(Z_3)+
     {(\theta_{23}D-\o\theta_{23}\o{D})\over Z_{23}}{\cal L}(Z_3) &(2.35)\cr
  &\quad\quad+ {\theta_{23}\o\theta_{23}\over Z_{23}}\partial_z
{\cal L}(Z_3)+reg.&\cr}$$
\meno
Furthermore, the definition (2.16) of super primary fields is equivalent
to the following OPE\meno
$$\eqalignno{ {\cal L}(Z_2)\>\Phi_{\Dt}^Q(Z_3) &=
     {\Dt\theta_{23}\o\theta_{23}\over Z_{23}^2}\Phi_{\Dt}^Q(Z_3)+
     {(\theta_{23}D-\o\theta_{23}\o{D})\over Z_{23}}\Phi_{\Dt}^Q(Z_3)
&(2.36)\cr
 &\quad\quad+ {\theta_{23}\o\theta_{23}\over Z_{23}}\partial_z
\Phi_{\Dt}^Q(Z_3)
    +{Q\over Z_{23}}\Phi(Z_3)+reg.   &\cr}$$
\meno
Here the coupling constants take the values $C_{{\cal L}\Phi\Phi}=Q/2$
and $\alpha_{{\cal L}\Phi\Phi}=2\Dt/Q$.
\pano
By expanding the fields in (2.29)(2.31)(2.33) into their components we are
able
to express the coupling constants of these components by the
superymmetric coupling constant. The long list of these
relations is given in appendix A..
Here we want to stress again that the explicit
construction of W-algebras has become tractable only with the
help of these relations which reduce the amount of
computer time and memory considerably. \bigno
2.5. SUPER NORMAL ORDERED PRODUCTS\meno
W-algebras are not Lie algebras in the classical sense because they contain
a multiplication of the generators. In the language of quantum field
theory, one says that a notion of a normal ordering operation is
necessary.
Since we want to use our formulae for $Osp(2|2)$ invariant fields
we have to define  super normal ordered products which are super
quasiprimary.
The way to proceed is analogous to the one in the N=0 and N=1 case. \pano
The fields $N_s(..)$ which occur in the regular part of the OPE can be
considered
as normal ordered products. The obstacle that they are not quasiprimary
can be overcome by projecting them onto fields ${\cal N}_s(..)$
with the desired property.
\pano
We consider the regular part of an OPE of two fields and define the
normal ordering operation $N_s(..)$ as follows.\meno
$$\eqalignno{ \Phi_i(Z_1)\Phi_j(Z_2)={\rm s.\>t. }
            &+\sum_{n=0}^{\infty}{Z_{12}^{n}\over n!}
             N_s(\Phi_j\partial_z^{n}\Phi_i)(Z_2)
            +\sum_{n=0}^{\infty}{\theta_{12}Z_{12}^{n}\over n!}
             N_s(\Phi_jD\partial_z^{n}\Phi_i)(Z_2)  &(2.37)\cr
           &+\sum_{n=0}^{\infty}{\o\theta_{12}Z_{12}^{n}\over n!}
             N_s(\Phi_j\o{D}\partial_z^{n}\Phi_i)(Z_2)
           +\sum_{n=0}^{\infty}{\theta_{12}\o\theta_{12}Z_{12}^{n}\over 2n!}
             N_s(\Phi_j[\o{D},D]\partial_z^{n}\Phi_i)(Z_2) &\cr }$$
\meno
The components of $N_s(..)$ can be obtained easily by a super Taylor
expansion of (2.37) (see appendix B).
Application of the derivatives $\partial_2$, $D_2$, $\o{D}_2$ and
$[\o{D}_2,D_2]$ to (2.37) gives the following properties of $N_s(..)$:
\meno
$$\eqalignno{ \partial N_s(\Phi_j\Phi_i)&=N_s(\Phi_j\partial\Phi_i)+
             N_(\partial\Phi_j\Phi_i) &\cr
        D N_s(\Phi_j\Phi_i)&=N_s(\Phi_j D\Phi_i)+(-1)^{2\Dt_i}
             N_s(D\Phi_j\Phi_i) &(2.38)\cr
        \o{D} N_s(\Phi_j\Phi_i)&=N_s(\Phi_j \o{D}\Phi_i)+(-1)^{2\Dt_i}
             N_s(\o{D}\Phi_j\Phi_i) &\cr
        [\o{D},D] N_s(\Phi_j\Phi_i)&=N_s(\Phi_j [\o{D},D]\Phi_i)-
             N_s([\o{D},D]\Phi_j\Phi_i) &\cr   }$$
\meno
We will not discuss
the procedure to get quasiprimary fields  in most
generality. But we will present some hopefully clarifying examples.
\pano
We introduce the quasiprimary projection
${\cal N}_s(\Phi_j\partial^n\Phi_i)$ in such a way that
$C_{ij}^k A_{ij(i+j+n)}^0={1\over n!}$ and
$C_{ij}^k E_{ij(i+j+n)}^{-1}=0$. The second condition is natural since
it gurantees that the field
${\cal N}_s(\Phi_j\partial^n\Phi_i)$ itself appears only once in
the OPE of $\Phi_i$ and $\Phi_j$.
These two conditions fix the two coupling constants for the quasiprimary
field and all derivatives couple due to (2.29),(2.30).
For the three other kinds of normal ordered products in (2.37)
the definition of
their quasiprimary projection is similar. In particular, for the the  fields
${\cal N}_s(\Phi_j[\o{D},D]\partial^n\Phi_i)$ we require
$C_{ij}^kA_{ij(i+j+n+1)}^0=0$ additionally. \pano
For explicit calculations one needs a representation of these $Osp(2|2)$
invariant normal ordered products in terms of the $N_s(..)$ fields and the
fields appearing in the singular part of the OPE. But it is already
sufficient
to know the lowest component since higher ones can be obtained
using (2.23).
\pano
We present some examples how this procedure works.
\meno
\hskip 1cm (i) ${\cal N}_s({\cal L}{\cal L})$ \smno
{}From the OPE \meno
$$ {\cal L}(Z_1){\cal L}(Z_2)={\rm sing.\>terms }+
    {\cal N}_s({\cal L}{\cal L})(Z_2)+{1\over 3}[\o{D},D]{\cal L}(Z_2)+...
\eqno(2.39)$$
\meno
one reads off directly \meno
$$ {\cal N}_s({\cal L}{\cal L})=N_s({\cal L}{\cal L})
           -{1\over 3}[\o{D},D]{\cal L}.\eqno(2.40) $$
\meno
The lowest component which must be $SU(1,1)$ invariant is\meno
$$ {\cal N}_s({\cal L}{\cal L})_1={\cal N}(JJ)
           -{2\over 3}L \eqno(2.41)$$
\meno
where ${\cal N}(JJ)$  denotes the $SU(1,1)$ invariant normal ordered
product. As expected from $Osp(2|2)$ invariance
$ {\cal N}_s({\cal L}{\cal L})_1 $ is orthogonal to $L$. \bigno
\hskip 1cm (ii) ${\cal N}_s({\cal L}[\o{D},D]{\cal L})$ \meno
$$ {\cal L}(Z_1){\cal L}(Z_2)={\ldots }+\theta_{12}\o\theta_{12}
   \left({1\over 2}{\cal N}_s({\cal L}[\o{D},D]{\cal L})(Z_2)+{3\over 20}
        [\o{D},D]{\cal N}_s({\cal L}{\cal L}) +O(\partial) \right)+\ldots
\eqno(2.42)$$
\meno
Now one calculates the lowest component and applies the $SU(1,1)$
projector.
\meno
$$ {\cal N}_s({\cal L}[\o{D},D]{\cal L})_1=2{\cal N}(JL)
           -{3\over 5}{\cal N}_s({\cal L}{\cal L})'_{\theta\o\theta}
            \eqno(2.43)$$
\meno
Note that the super fields $ {\cal N}_s({\cal L}D{\cal L}) $,
$ {\cal N}_s({\cal L}\o{D}{\cal L}) $ vanish. Otherwise they would
contribute in (2.42) as well.\meno
\hskip 1cm (iii) ${\cal N}_s({\cal L}D\partial{\cal L})$ \meno
$$ {\cal L}(Z_1){\cal L}(Z_2)={\ldots }+\theta_{12}
   \left({\cal N}_s({\cal L}D\partial{\cal L})(Z_2)+{1\over 6}
        D{\cal N}_s({\cal L}[\o{D},D]{\cal L}) +O(\partial) \right)+\ldots
\eqno(2.44)$$
\meno\vfill\eject\pano
Hence\meno
$$ {\cal N}_s({\cal L}D{\cal L})_1={1\over \sqrt 2}{\cal N}(J\partial\oG)
          -{1\over \sqrt 2}{1\over 6}
        {\cal N}_s({\cal L}[\o{D},D]{\cal L})_{\o\theta}.\eqno(2.45)$$
\meno
With this procedure we are able to generate higher normal
ordered products successively. \pano
Applying the $Osp(2|2)$ projector to (2.38) yields the following properties
of the super quasiprimary fields\meno
$$\eqalignno{ {\cal N}_s(\Phi_j\partial\Phi_i)&=-
             {\cal N}_s(\partial\Phi_j\Phi_i) &\cr
        {\cal N}_s(\Phi_j D\Phi_i)&=(-1)^{2\Dt_i+1}
             {\cal N}_s(D\Phi_j\Phi_i) &(2.46)\cr
        {\cal N}_s(\Phi_j \o{D}\Phi_i)&=(-1)^{2\Dt_i+1}
             {\cal N}_s(\o{D}\Phi_j\Phi_i) &\cr
        {\cal N}_s(\Phi_j [\o{D},D]\Phi_i)&=
             {\cal N}_s([\o{D},D]\Phi_j\Phi_i). &\cr   }$$
\meno
Furthermore, the operation ${\cal N}_s(..)$ is antisymmetric for superfields
of half-integer dimension and symmetric otherwise. This property together
with (2.46) implies for example\meno
$$ {\cal N}_s({\cal L}{D}{\cal L})={\cal N}_s({\cal L}\o{D}{\cal L})=0.
\eqno(2.47)$$
\meno
Now we have developped the N=2 supersymmetric structure which is necessary
for the explicit construction of N=2 super W-algebras. We have defined
quasiprimary normal ordered products so that we are allowed to apply
the formulas (2.26)-(2.34) to them. In the next section we explicitly
construct
examples of N=2 super W-algebras with one additional
pair of super primary fields besides the Virasoro field.
\vfill\eject
\smno
{\cl{\bf 3. N=2 super W-algebras}}
\bigno
In this section we apply the formalism of the last section to the
explicit construction of N=2 super W-algebras. We restrict ourselves to
extensions of the N=2 super Virasoro algebra by one single
pair of super primary fields.
Note that it would be hardly a feasible task to
explicitly construct a general N=0 W-algebra containing twelve generators.
But the N=2 structure is so restrictive that such a complex calculation
becomes practicable. Up to now one knows only two nonlinear quantum
N=2 super W-algebras explicitly. These are the  ${\cal SW}(1,3/2)$ [13] and
the ${\cal SW}(1,2)$ [14] algebra. The first one has been studied by Inami
et al. using an N=2 conformal bootstrap algorithm. Our results show
that there are more restrictions on the allowed $c$ values than given
in [13], where this, however, had already been expected.
The second algebra has been calculated
by Romans using algebraic methods. But he considered this algebra only
for vanishing $U(1)$ charge $Q$ of the spin two generator.
For this case he found that the ${\cal SW}(1,2)$ algebra exists for generic
value of the
central charge $c$. \pano
In this paper we extend this series up to spin $5/2$ and we allow
the additional generators to have non-vanishing charge $Q$.
For ${\cal SW}(1,3)$ we give a maximal set of
values of $c$ and $Q$ for which this algebra may exist.\bigno
3.1. ALGORITHM\meno
We want to extend the N=2 super Virasoro algebra by a pair of super primary
fields $\Phi_{\Dt}^Q$ and $\Phi_{\Dt}^{-Q}$. For generic charge $Q$ the
OPEs $\Phi_{\Dt}^{\pm Q}(Z_1)\>\Phi_{\Dt}^{\pm Q}(Z_2)$ are trivial and
the OPE $\Phi_{\Dt}^{+Q}(Z_1)\>\Phi_{\Dt}^{-Q}(Z_2)$ contains in its
singular part only descendants of the super Virasoro field ${\cal L}$.
For integer dimension $\Dt$ a self coupling is only possible if the charge
is zero, for half-integer dimension one needs charge $Q=\pm 1$.
\pano
We choose the following normalisation of the two point function
\meno
$$\la\Phi_{\Dt}^{+Q}(Z_1)\>\Phi_{\Dt}^{-Q}(Z_2)\ra={{c/\Dt} \over
    Z_{12}^{2\Dt}}\left( 1-{Q\over 2}{\theta_{12}\o\theta_{12}\over Z_{12}
    }\right).\eqno(3.1) $$
\meno
For explicit computer calculations we expand the components into
Fourier-modes making it
possible to apply Lie algebra methods like for example the universal
polynomials in (2.4)-(2.7). For consistency of the algebra, all commutators
have to satisfy Jacobi identities. Checking this leads to restrictions
on the central charge $c$ and the $U(1)$ charge $Q$.
We should note that an important advantage of the mode approach compared
to OPE's is that it allows more directly to
investigate highest weight representations [28].
Thus, we proceed as follows:\smno
\hskip 1cm (a) We write down the most general ansatz for the super OPE.
To this end, we need a basis of super quasiprimary normal ordered
products for each occuring dimension. As already mentioned it is
sufficient to know the lowest components of these fields.\smno
\hskip 1cm (b) Then we expand the super fields into their components
and calculate the structure constants for the lowest components. The
structure constants for the higher components are given by the formulae
presented in appendix A.
Note that N=2 supersymmetry implies that we only have to calculate
one $D_{\delta}$-matrix for every super conformal dimension $\Dt=\delta$.
Thus the effort minimizes considerably.\smno
\hskip 1cm (c) Finally, we have to check Jacobi identities. It
is sufficient to check them only for the additional super primaries because
Jacobi identities including the super Virasoro field are satisfied
automatically by $Osp(2|2)$ invariance. Furthermore, some Jacobi identities
are related by charge conjugation $Q\to -Q$. \pano We denote by
$(\varphi_i\>\varphi_j\>\varphi_k\>\varphi_l)$ the factor of
the Jacobi identity
$[\varphi_i,[\varphi_j,\varphi_k]]+cycl.$ in front of the field
$\varphi_l^*$.
$\varphi_l^*$ is the adjoint of $\varphi_l$ with respect to the form
$\la .,.\ra$, for example $(\psi^+)^*=\o\psi^-$.
Writing down the Jacobi identities explicitly
one obtains:\meno
If $\delta_i=\delta_j$ and $\delta_k=\delta_l$ then
$$(\varphi_i\>\varphi_j\>\varphi_k\>\varphi_l)=0\>\Rightarrow\>
(\varphi_j\>\varphi_i\>\varphi_l\>\varphi_k)=0. \eqno(3.2)$$
\meno
Exploiting these symmetries, one has to check only the following
twelve Jacobi identities\meno
$$\eqalignno{
&(\varphi^+\>\varphi^+\>\varphi^-\>\varphi^-)\quad\quad
      (\chi'^+\>\chi'^+\>\chi'^-\>\chi'^-)  &\cr
&(\varphi^+\>\varphi^+\>\varphi^-\>\chi'^-)\quad\quad
(\varphi^+\>\chi'^+\>\chi'^-\>\chi'^-) &\cr
&(\varphi^+\>\varphi^+\>\chi'^-\>\chi'^-)\quad\quad
(\varphi^+\>\varphi^-\>\chi'^+\>\chi'^-) &(3.3)\cr
&(\varphi^+\>\varphi^+\>\psi^-\>\o\psi^-)\quad\quad
  (\varphi^+\>\varphi^-\>\psi^+\>\o\psi^-) &\cr
&(\psi^+\>\o\psi^-\>\chi'^+\>\chi'^-)\quad\quad
 (\psi^+\>\o\psi^+\>\chi'^-\>\chi'^-) &\cr
&(\psi^+\>\o\psi^-\>\psi^+\>\o\psi^-)\quad\quad
 (\psi^+\>\o\psi^+\>\psi^-\>\o\psi^-). &\cr }$$
\meno
Our calculations assure us that N=2 supersymmetry generates
further relations among the Jacobi identities above.
\bigno
3.2. {\cal SW}(1,$\Dt$) ALGEBRAS
\meno
In this subsection we present the results of our calculations.
We should note beforehand that all the  N=2 super W-algebras considered
only exist for some discrete
values of the central charge $c$ and the $U(1)$ charge $Q$.
\meno
{\bf {\cal SW}(1,3/2)}
\smno
In the OPE
 $\Phi_{3/2}^{+Q}(Z_1)\>\Phi_{3/2}^{-Q}(Z_2)$ the following
super quasiprimary fields may occur for generic value of $Q$.
\bigno
$$\vbox{\offinterlineskip\tabskip=0pt
\halign{ \strut \vrule #& \quad $#$ \quad \hfil &\vrule #&\quad $#$  \hfil
   \quad & \vrule #& \quad $#$  \hfil\quad &\vrule #\cr
\noalign{\hrule}
&\Dt && Q && fields & \cr
\noalign{\hrule}
&1 && 0 && {\cal L} & \cr
&2 && 0 && {\cal N}_s({\cal L}{\cal L}) &\cr
&3 && 0 && {\cal N}_s({\cal N}_s({\cal L}{\cal L}){\cal L}) &\cr
&3 && 0 && {\cal N}_s({\cal L}[\o{D},D]{\cal L}) &\cr
\noalign{\hrule}
\noalign{\smallskip}
\multispan 7 {\klein table 1 }\hfill \cr }}$$
\bigno
where the first component of the field
${\cal N}_s({\cal N}_s({\cal L}{\cal L}){\cal L})$ is given by
\meno
$${\cal N}_s({\cal N}_s({\cal L}{\cal L}){\cal L})_1=
   {\cal N}({\cal N}_s({\cal L}{\cal L})_1 J)-{2\over 5}
   {\cal N}_s({\cal L}{\cal L})'_{\theta\o\theta}.  \eqno(3.4)$$
\meno
The Kac-determinants in the vacuum sector are
\bigno
$$\vbox{\offinterlineskip\tabskip=0pt
\halign{ \strut \vrule #& \quad $#$ \quad &\vrule #&\quad $#$  \hfil
\quad &\vrule # \cr
\noalign{\hrule}
&\Delta && det(D_{\Delta})\sim  &\cr
\noalign{\hrule}
&1 && c & \cr
&2 && c(c-1)& \cr
&3 && c^2(2c-3)(c+6)(c-1) &\cr
\noalign{\hrule}
\noalign{\smallskip}
\multispan 5 {\klein table 2 }\hfill \cr }}$$
\bigno
The structure constants $C_{ij}^k$, $\alpha_{ijk}$ for the normal
ordered products are rational functions in $c$ and $Q$: \meno
$$\eqalignno{
C_{\Phi^+\Phi^-}^{\cal L}&=Q \quad\quad\quad\quad \alpha={3\over Q} &\crns
C_{\Phi^+\Phi^-}^{{\cal N}({\cal L}{\cal L})}&={3(Q^2-1)\over 2(c-1)}
    \quad\quad \alpha={8Q\over 3(Q^2-1)} &\cr & &(3.5)\cr
C_{\Phi^+\Phi^-}^{{\cal N}({\cal L}[\o{D},D]{\cal L})}(\alpha-1)&=-
{(3c-2Q-4)(Q^2-9)\over 3(2c-3)(c+6)} &\crns
C_{\Phi^+\Phi^-}^{ {\cal N}({\cal N}({\cal L}{\cal L}){\cal L})}
(\alpha-1)&=-{(2c(Q^2-1)+13Q^2-1)(Q^2-9)\over 2(2c-3)(c+6)(c-1)} &\cr }$$
\meno
In the following table we list the rational values of the parameters $c$
and $Q$
for which all Jacobi identities (3.3) hold.
\bigno
$$\vbox{\offinterlineskip\tabskip=0pt
\halign{ \strut \vrule #& \hfil \quad $#$ \quad &\vrule #&\quad $#$  \hfil
\quad &\vrule # \cr
\noalign{\hrule}
&Q && c & \cr
\noalign{\hrule}
&0 && 9/4, 5/3, -3/2 & \cr
&\pm 3 && 9 &\cr
&\pm 1 && 3 &\cr
\noalign{\hrule}
\noalign{\smallskip}
\multispan 5 {\klein table 3 }\hfill \cr }}$$
\bigno
For $\Dt=\pm Q/2$ the fields $\Phi_{\Dt}^{Q}$ and $\Phi_{\Dt}^{-Q}$
are so-called primary (anti)chiral superfields [19], i.e. they satisfy
$\o{D}\Phi_{\Dt}^Q=0$ for the chiral, and
$D\Phi_{\Dt}^Q=0$ for the antichiral case. The chiral ${\cal SW}(1,3/2)$
algebra
had already been constructed in ref. [13,20]. It is of particular
interest because it exists for $c=9$ which allows it to be used
4D string compactification.
One important drawback, however, is that according to an argument of
Cardy [21,22]
no minimal representations are allowed to exist
for an algebra with four bosonic and four fermionic generators for
$c\ge 6$.
For a general $SW(1,\Dt)$ algebra with twelve generators the same
argument implies that only for $c<9$ minimal representations are
possible.\pano
We remark that by constructing this algebra using Jacobi identities
we obtain only a subset of the rational $c$ values claimed by Inami et al..
They found that the four-point functions of the components are associative
for generic value of $c$. This behaviour is very strange since in
the N=0 and N=1 case the two requirements are equivalent.
\smno
For the charge $Q=\pm 1$ the fields $\Phi_{\Dt}^{1}$ and $\Phi_{\Dt}^{-1}$
themselves could occur in all OPEs of the two super primary fields.
This implies the appearance of the following descendants
$$\vbox{\offinterlineskip\tabskip=0pt
\halign{ \strut \vrule #& \quad $#$ \quad \hfil &\vrule #&\quad \hfil $#$
   \quad & \vrule #& \quad $#$  \hfil\quad &\vrule #\cr
\noalign{\hrule}
&\Dt && Q && fields  &\cr
\noalign{\hrule}
&5/2 && +1 &&{\cal N}_s(\Phi^{+1}{\cal L}) & \cr
&5/2 && -1 &&{\cal N}_s(\Phi^{-1}{\cal L}) & \cr
&3 && 0 &&{\cal N}_s(\Phi^{1}D{\cal L}) & \cr
&3 && 0 &&{\cal N}_s(\Phi^{-1}\o{D}{\cal L})  &\cr
&3 && 2 &&{\cal N}_s(\Phi^{1}\o{D}{\cal L}) & \cr
&3 && -2 &&{\cal N}_s(\Phi^{-1}{D}{\cal L}) & \cr
\noalign{\hrule}
\noalign{\smallskip}
\multispan 7 {\klein table 4 }\hfill \cr }}$$
\bigno
Realize that the last two fields may only occur in the OPEs
$\Phi_{3/2}^{+1}(Z_1)\>\Phi_{3/2}^{+1}(Z_2)$ and
$\Phi_{3/2}^{-1}(Z_1)\>\Phi_{3/2}^{-1}(Z_2)$, respectively.
Repeating the whole calculation with this further extended algebra,
we obtain that all Jacobi identities are satisfied only for the
central charges $c=3$ and $c=-1$. For $c=-1$ we obtain
the additional consistency condition \meno
$$C_{\Phi^{+1}\Phi^{-1}}^{\Phi^{+1}}
C_{\Phi^{+1}\Phi^{-1}}^{\Phi^{-1}}=-{128\over 15} .\eqno(3.6)$$
\bigno
{\bf {\cal SW}(1,2)}
\smno
In the OPE super quasiprimary fields up to dimension four may occur.
The new ones are listed in the following table
\bigno
$$\vbox{\offinterlineskip\tabskip=0pt
\halign{ \strut \vrule #& \quad $#$ \quad \hfil &\vrule #&\quad \hfil $#$
   \quad & \vrule #& \quad $#$  \hfil\quad &\vrule #\cr
\noalign{\hrule}
&\Dt && Q && fields & \cr
\noalign{\hrule}
&7/2 && -1 && {\cal N}_s({\cal L}{D}\partial{\cal L}) &\cr
&7/2 && 1 && {\cal N}_s({\cal L}\o{D}\partial{\cal L}) &\cr
&4 && 0 && {\cal N}_s({\cal N}_s({\cal N}_s({\cal L}{\cal L})
               {\cal L}){\cal L}) &\cr
&4  && 0 && {\cal N}_s({\cal N}_s({\cal L}[\o{D},D]{\cal L}){\cal L}) &\cr
&4 && 0 && {\cal N}_s({\cal L}\partial^2{\cal L}) &\cr
\noalign{\hrule}
\noalign{\smallskip}
\multispan 7 {\klein table 5 }\hfill \cr }}$$
\bigno
The new determinants of the $D_{\Delta}$ matrices are
\bigno
$$\vbox{\offinterlineskip\tabskip=0pt
\halign{ \strut \vrule #& \quad $#$ \quad &\vrule #&\quad $#$  \hfil
\quad &\vrule # \cr
\noalign{\hrule}
&\Delta && det(D_{\Delta})\sim  &\cr
\noalign{\hrule}
&7/2 && c^2(c-1)^2 & \cr
&4 && c^3(5c-9)(2c-3)(c+6)(c+1)(c-1)^2& \cr
\noalign{\hrule}
\noalign{\smallskip}
\multispan 5 {\klein table 6 }\hfill \cr }}$$
\bigno
This algebra exists for the following $c$ and $Q$ values
\bigno
$$\vbox{\offinterlineskip\tabskip=0pt
\halign{ \strut \vrule #& \quad \hfil $#$ \quad &\vrule #&\quad $#$  \hfil
\quad &\vrule # \cr
\noalign{\hrule}
&Q && c & \cr
\noalign{\hrule}
&0 && 12/5, -3 & \cr
&\pm 4 && 12, -9, -21,  &\cr
&\pm 3/2 && 15/4 &\cr
\noalign{\hrule}
\noalign{\smallskip}
\multispan 5 {\klein table 7 }\hfill \cr }}$$
\bigno
Our results are in agreement with those of Romans [14]. The values
$c=12/5$ and $c=-3$ are just the zeros of the self-coupling
$C_{\Phi\Phi}^{\Phi}$. Besides the chiral case $c=12$,
the value $c=15/4$ is very remarkable because it is greater than three.
In the N=0 and N=1 case one has never
found a similar behaviour: There, all the
W-algebras, which do not exist generically have central charges $c<1$ or
$c<3/2$, respectively.
In the next section we will conjecture that this theory could be related
to the non-compact coset model $(SU(1,1)/U(1))\times U(1)$.
\bigno
{\bf {\cal SW}(1,5/2)}
\smno
In the OPE super quasiprimary fields up to dimension five may occur.
\bigno
$$\vbox{\offinterlineskip\tabskip=0pt
\halign{ \strut \vrule #& \quad $#$ \quad \hfil &\vrule #&\quad\hfil  $#$
   \quad & \vrule #& \quad $#$  \hfil\quad &\vrule #\cr
\noalign{\hrule}
&\Dt && Q && fields & \cr
\noalign{\hrule}
&9/2 && -1 && {\cal N}_s({\cal N}_s({\cal L}{D}\partial{\cal L}){\cal L})
&\cr
&9/2 && 1 && {\cal N}_s({\cal N}_s({\cal L}\o{D}\partial{\cal L}){\cal L})
&\cr
&5 && 0 && {\cal N}_s({\cal N}_s({\cal N}_s({\cal N}_s({\cal L}{\cal L})
               {\cal L}){\cal L}){\cal L}) &\cr
&5  && 0 && {\cal N}_s({\cal N}_s({\cal N}_s({\cal L}[\o{D},D]{\cal L})
          {\cal L}){\cal L}) &\cr
&5 && 0 && {\cal N}_s({\cal N}_s({\cal L}\partial^2{\cal L}){\cal L}) &\cr
&5 && 0 && {\cal N}_s({\cal N}_s({\cal L}\o{D}\partial{\cal L})D{\cal L})
&\cr
&5 && 0 && {\cal N}_s({\cal N}_s({\cal L}D\partial{\cal L})\o{D}{\cal L})
&\cr
&5 && 0 && {\cal N}_s({\cal L}[\o{D},D]\partial^2{\cal L}) &\cr
\noalign{\hrule}
\noalign{\smallskip}
\multispan 7 {\klein table 8 }\hfill \cr }}$$
\bigno
The Kac determinants of the $D_{\Delta}$ matrices are
\bigno
$$\vbox{\offinterlineskip\tabskip=0pt
\halign{ \strut \vrule #& \quad $#$ \quad &\vrule #&\quad $#$  \hfil
\quad &\vrule # \cr
\noalign{\hrule}
&\Delta && det(D_{\Delta})\sim  &\cr
\noalign{\hrule}
&9/2 && c^2(c-1)^2(2c-3)^2 & \cr
&5 && c^6(5c-9)(2c-3)^2 (c+12)(c+6)^3(c+1)(c-1)^5(c-2)& \cr
\noalign{\hrule}
\noalign{\smallskip}
\multispan 5 {\klein table 9 }\hfill \cr }}$$
\bigno
This algebra exists for the following $c$ and $Q$ values
\bigno
$$\vbox{\offinterlineskip\tabskip=0pt
\halign{ \strut \vrule #& \quad\hfil $#$ \quad &\vrule #&\quad $#$  \hfil
\quad &\vrule # \cr
\noalign{\hrule}
&Q && c & \cr
\noalign{\hrule}
&0 && 5/2, 9/7, -9/2 & \cr
&\pm 5 && 15 &\cr
&\pm 2 && 9/2 &\cr
&\pm 1 && 3 &\cr
\noalign{\hrule}
\noalign{\smallskip}
\multispan 5 {\klein table 10}\hfill \cr }}$$
\bigno
Just like the ${\cal SW}(1,3/2)$ algebra, this algebra exists for $Q=\pm 1$
and $c=3$.
\bigno
{\bf {\cal SW}(1,3)}
\smno
Since the commutator of the lowest components
contains only fields with maximal dimension five, we could calculate
the Jacobi identity
$(\varphi^+\>\varphi^+\>\varphi^-\>\varphi^-)$ without generating new
normal ordered products. For vanishing charge $Q$ the Jacobi identity
is satisfied trivially. But for non-vanishing charge one gets two
conditions which have only finitly many rational solutions for $Q$ and
$c$.
\bigno
$$\vbox{\offinterlineskip\tabskip=0pt
\halign{ \strut \vrule #&\hfil \quad $#$ \quad &\vrule #&\quad $#$  \hfil
\quad &\vrule # \cr
\noalign{\hrule}
&Q && c & \cr
\noalign{\hrule}
&0 && ? & \cr
&\pm 6 && 18, -15, -33 &\cr
&\pm 5/2 && 21/4 &\cr
&\pm 4/3 && 10/3 &\cr
&\pm 3/4 && 45/16 &\cr
\noalign{\hrule}
\noalign{\smallskip}
\multispan 5 {\klein table 11}\hfill \cr }}$$
\bigno
Comparing with the ${\cal SW}(1,2)$ algebra, we suppose that the last two
solutions
will turn out to be inconsistent with the remaining Jacobi identities.
\bigno
\cl{{\bf 4. Discussion }}
\meno
In this section we shall speculate on the general structure of
N=2 ${\cal SW}(1,\Dt)$ algebras regarding
the data presented in the previous section.
For the time being, we can only extract a few hints
for the complete classification of these algebras.\pano
In the N=0 and N=1 case, the discrete values of $c$ for which W-algebras
exist can be understood from the representation theory of the (super)
Virasoro algebra. In the N=2 case the situation is similar. But here
also the unitary non-minimal representations play a r\^ole.
Let us bring some order into the list of existing N=2 super W-algebras.
\meno
\hskip 1cm (a)
The following table displays the $c$-values contained in the unitary series
(2.15) of the N=2 super Virasoro algebra
\bigno
$$\vbox{\offinterlineskip\tabskip=0pt
\halign{ \strut \vrule #& \quad $#$ \quad \hfil &\vrule #&\quad\hfil  $#$
   \quad & \vrule #& \quad $#$  \hfil\quad &\vrule #
   & \quad $#$  \hfil\quad &\vrule #
   & \quad $#$  \hfil\quad &\vrule #\cr
\noalign{\hrule}
&\Dt && Q && c && k=l && m &\cr
\noalign{\hrule}
&3/2 && 0 && 9/4 && 6 && 0 &\cr
&2 && 0 && 12/5 && 8 && 0 &\cr
&5/2 && 0 && 5/2 && 10 && 0 &\cr
\noalign{\hrule}
\noalign{\smallskip}
\multispan {11}{\klein table 12}\hfill \cr }}$$
\bigno
where $k,l,m$ label the minimal unitary representations of the N=2 super
Virasoro algebra in the Neveu-Schwarz sector [5]
\meno
$$\vbox{
\halign{  \quad ${\ds #}$  \hfil &\quad\quad $#$ \hfil
   & $\>$ $#$  \hfil \cr
c={3k\over k+2}&  k=1,2... &\crnm
\Dt= {l(l+2)-m^2\over 4(k+2)}&  l=0,...,k &\crnm
Q= {m\over (k+2)} & m=-l,-l+2,...,l &\cr }}$$
$$\eqno(4.1)$$
\meno
Calculation of the fusion rule [18] for the field $\Phi_{l=k}^{m=0}$ shows
that it is indeed a simple current [23]
\meno
$$[\Phi_{l=k}^{m=0}]\times[\Phi_{l=k}^{m=0}]=[1]. \eqno(4.2)$$
\meno
For integer dimension $k=4\Dt$ the list of known modular invariant
partition functions [24]
contains a suitable, so called integer spin, off-diagonal invariant which
can be interpreted as the diagonal invariant of the extended algebra
${\cal SW}(1,\Dt)$.
For half-integer dimension, i.e. $k=4\Dt=4j-2$, there exists a
so called automorphism invariant $D_{2j+1}$
\meno
$$\sum_{l=0}^{k/2}|\sigma_{2l}|^2+\sum_{l=0}^{2j-1}\sigma_{2l+1}
      \sigma_{k-2l-1}^* \eqno(4.3) $$
\meno
up to a diagonal $U(1)$ part. \pano
This partition function can be rewritten in a fermionic form
\meno
$$\eqalignno{ {1\over 2}\sum_{l=0}^{j-1} \big( |\sigma_{2l}
            +\sigma_{k-2l}|^2
           +|\sigma_{2l}&-\sigma_{k-2l}|^2  \big) &(4.4)\cr
       &+ {1\over 2}\sum_{l=0}^{j-1}\left( |\sigma_{2l+1}
        +\sigma_{k-2l-1}|^2
      -|\sigma_{2l+1}-\sigma_{k-2l-1}|^2 \right) &\cr }  $$
\meno
where the first two terms form the Neveu-Schwarz sector and the second
two the Ramond sector. The last term is the
Witten-index of the representations of the extended algebra and therefore
a constant.
Thus, omitting it leads to a diagonal fermionic
partition function for the ${\cal SW}(1,\Dt)$ algebra.
\meno\vfill\eject\pano
\hskip 1cm (b)
For $Q=0$ one can conjecture two other series
\bigno
$$\vbox{\offinterlineskip\tabskip=0pt
\halign{ \strut \vrule #& \quad $#$ \quad \hfil &\vrule #&\quad\hfil  $#$
   \quad & \vrule #& \quad $#$  \hfil\quad &\vrule #  \cr
\noalign{\hrule}
&\Dt && c && k &\cr
\noalign{\hrule}
&3/2 && 5/3 && 5/2 &\cr
&5/2 && 9/7 && 3/2 &\cr
\noalign{\hrule}
\noalign{\smallskip}
\multispan 7{\klein table 13}\hfill \cr }}$$
\meno
Note that the half-integer values of $k$ give no minimal representations
of the N=2 super Virasoro algebra.
It is not clear how this series will continue.
There also occurs a non-unitary series with negative values of $c$.
\bigno
$$\vbox{\offinterlineskip\tabskip=0pt
\halign{ \strut \vrule #& \quad $#$ \quad \hfil &\vrule #&\quad\hfil  $#$
   \quad & \vrule #\cr
\noalign{\hrule}
&\Dt && c  &\cr
\noalign{\hrule}
&3/2 && -3/2  &\cr
&2 && -3  &\cr
&5/2 && -9/2  &\cr
\noalign{\hrule}
\noalign{\smallskip}
\multispan 5{\klein table 14}\hfill \cr }}$$
\meno
These values fit nicely into $c=3(1-\Dt)$ and we
conjecture the existence of all ${\cal SW}(1,\Dt)$ algebras for
this central charge in analogy to the descendant series of negative
$c$ values in the N=0 and N=1 case. In ref. [25] these W-algebras have
been investigated in more detail presenting a vertex operator
construction of the generators.
\meno
\hskip 1cm (c)
For $Q\ne 0$ the spectral flow algebras occur.
\bigno
$$\vbox{\offinterlineskip\tabskip=0pt
\halign{ \strut \vrule #& \quad $#$ \quad \hfil &\vrule #&\quad\hfil  $#$
   \quad & \vrule #& \quad $#$  \hfil\quad &\vrule #  \cr
\noalign{\hrule}
&\Dt && Q && c &\cr
\noalign{\hrule}
&3/2 && \pm 3 && 9 &\cr
&2 && \pm4 && 12 &\cr
&5/2 && \pm 5 && 15 &\cr
&3 && \pm6 && 18 &\cr
\noalign{\hrule}
\noalign{\smallskip}
\multispan 7{\klein table 15}\hfill \cr }}$$
\meno
It is well known that the N=2 Virasoro algebra has a non-trivial
automorphism which continuously connects the Neveu-Schwarz and the
Ramond sector [26]. It acts in the following way
\meno
$$\eqalignno{
J_n&\to J_n+{c\over 3}\alpha\delta_{n,0}&\cr
G^{\pm}_n&\to G^{\pm}_{n\pm\alpha}&(4.5)\cr
L_n&\to J_n+\alpha J_n+{c\over 6}\alpha^2\delta_{n,0}.&\cr }$$
\meno
For $\alpha=\pm 1$ one obtains a (anti)chiral spectral flow operator
$\Phi_{c/6}^{\pm c/3}$. The extension of the N=2 Virasoro algebra
by these spectral flow operators yields the W-algebras above.
The representation theory of these algebras has been studied by
Odake [27,28] confirming that no minimal representations exist.\pano
Note that for integer dimension $\Dt$ the chiral algebras exist also
for two additional negative values of $c$.
\bigno
$$\vbox{\offinterlineskip\tabskip=0pt
\halign{ \strut \vrule #& \quad $#$ \quad \hfil &\vrule #&\quad\hfil  $#$
   \quad & \vrule #& \quad $#$  \hfil\quad &\vrule #  \cr
\noalign{\hrule}
&\Dt && Q && c &\cr
\noalign{\hrule}
&2 && \pm4 && -9,-21 &\cr
&3 && \pm6 && -15,-33 &\cr
\noalign{\hrule}
\noalign{\smallskip}
\multispan 7{\klein table 16}\hfill \cr }}$$
\bigno
Assuming that $c$ depends only linearly on $\Dt$ the values above fit
into the two series $c=3(1-2\Dt)$ and $c=3(1-4\Dt)$.
\meno
\hskip 1cm (d)
The unitary necessarily non-unitary representations with $c>3$ have
been constructed by Dixon et al.
[29] from the non-compact coset model  $(SU(1,1)/U(1))\times
U(1)$.
The following W-algebras should be connected to this coset.
\bigno
$$\vbox{\offinterlineskip\tabskip=0pt
\halign{ \strut \vrule #& \quad $#$ \quad \hfil &\vrule #&\quad\hfil  $#$
   \quad & \vrule #& \quad $#$  \hfil\quad &\vrule #
   & \quad $#$  \hfil\quad &\vrule #
   & \quad $#$  \hfil\quad &\vrule #
   & \quad $#$  \hfil\quad &\vrule #\cr
\noalign{\hrule}
& \Dt && Q && c && k && j && m &\cr
\noalign{\hrule}
& 2 && \pm 3/2 && 15/4 && 10 && 5 && \mp6 &\cr
& 5/2 && \pm 2 && 9/2 && 6 && 3 && \mp4 &\cr
& 3 && \pm 5/2 && 21/4 && 14/3 && 7/3 && \mp10/3 &\cr
\noalign{\hrule}
\noalign{\smallskip}
\multispan {13}{\klein table 17}\hfill \cr }}$$
\bigno
with
$$\vbox{
\halign{  \quad ${\ds #}$  \hfil &\quad\quad $#$ \hfil
   & $\>$ $#$  \hfil \cr
c={3k\over k-2}& k>2&\crnm
\Dt= {-j(j-1)-m^2\over (k+2)}& j=n+\epsilon& \epsilon=0,1/2 \cr
&j=n-\epsilon& \epsilon=0,1/2 \cr
Q= -{2m\over (k-2)}& m=j+r& r\in Z_0^+ \cr
& m=-j-r& r\in Z_0^+ \cr }}$$
$$\eqno(4.6)$$
\meno
Besides the mentioned positive series $D_{n}^{+}$ and negative
series $D_{n}^{-}$, there also exist
a continuous series which, however, is not relevant for
our purpose.
Thus, we conjecture a series of W-algebras ${\cal SW}(1,\Dt)$ to exist for
\meno
$$c={3\over 4}(2\Dt+1)\quad\quad\quad Q=\pm(\Dt-{1\over 2}) \eqno(4.7)$$
\meno
In particlar, the ${\cal SW}(1,11/2)$ should  exist for $c=9$ and
$Q=\pm 5$. \pano
Note that the $c$-values in table 17 are also contained in the unitary
series of the Kazama-Suzuki model $SU(3)/(SU(2)\times U(1))$, which
is probably related to the ${\cal SW}(1,2)$ algebra of Romans with
\meno
$$c={6k\over k+3},\quad\quad\quad k\in \BZ_+\eqno(4.8)$$
\meno
For $\Dt>7/2$ we suppose a connection to the non-compact coset model
$SU(2,1)/(SU(2)\times U(1))$ of Bars [30], which has the following
unitary series
\meno
$$c={6k\over k-3}\quad\quad\quad k>3\eqno(4.9)$$
\meno
It would be interesting to clarify the connections among compact or
non-compact Ka\-za\-ma-Suzuki models, Romans' N=2 super $W_3$ algebra and
our ${\cal SW}(1,\Dt)$ series.
\meno
\hskip 1cm (e)
Our last conjecture reads that for half-integer dimension the W-algebra
exists for $c=3$ and $Q=\pm 1$.
\bigno
$$\vbox{\offinterlineskip\tabskip=0pt
\halign{ \strut \vrule #& \quad $#$ \quad \hfil &\vrule #&\quad\hfil  $#$
   \quad & \vrule #& \quad $#$  \hfil\quad &\vrule #  \cr
\noalign{\hrule}
&\Dt && Q && c &\cr
\noalign{\hrule}
&3/2 && \pm 1 && 3 &\cr
&5/2 && \pm 1 && 3 &\cr
\noalign{\hrule}
\noalign{\smallskip}
\multispan 7{\klein table 18}\hfill \cr }}$$
\bigno
We expect that there exists a free field construction of these
W-algebras with one chiral and one antichiral fermion.
\smno
\bigno
{\cl{\bf 5. Summary}}
\meno
In this paper we have investigated
extensions of the N=2 super Virasoro algebra, using an explicit N=2
supersymmetric formalism.
The motivation for this project was to get some insight into the
structure of N=2 symmetry algebras which do not exist for generic value
of $c$, in contrast to the
Kazami-Suzuki models. There also is some hope to
find a striking W-algebra as a new candidate for the string vacuum
in a Gepner-like construction. \pano
We have found that the structure of the simplest N=2 extended algebras, the
${\cal SW}(1,\Dt)$ series, is different from the analogous N=0 and N=1 cases.
The most interesting property --- and very important for applications
in string theory ---
is that these algebras are allowed to exist for $c\ge 3$.
Starting from the examples calculated we have conjectured the existence of
series of ${\cal SW}(1,\Dt)$
algebras. A proof would have to use
other methods like free field constructions and
vertex operators. \pano
As for the value $c=9$ distinguished in string theory, we conjecture
the ${\cal SW}(1,11/2)$ to exist for $Q=\pm 5$ besides the spectral flow
algebra ${\cal SW}(1,3/2)$ with $Q=\pm 3$, which is already known.
Unfortunately, these two algebras
do not admit minimal representations for $c=9$. To obtain minimal
representations, one has to investigate N=2 W-algebras with more
generators. Nevertheless, the ${\cal SW}(1,\Dt)$ algebras provide some
new examples of extended N=2 symmetries.
\meno
It is a pleasure to thank W. Nahm, W. Eholzer, M. Flohr,
R. H\"ubel,
J. Kellendonk, S. Mallwitz, M. R\"osgen,
and R. Varnhagen for discussion and especially A. Honecker,
A. Recknagel and M. Terhoeven for
reading the manuscript.\meno
\bigno
{\cl{\bf Appendix A}}\meno
In this appendix we express the component structure
constants by the supersymmetric coupling constants. We use the
abreviation $\Dt_{ijk}=\Dt_i+\Dt_j-\Dt_k$.
\pano
Neglecting $1/\sqrt 2$ factors we consider
superfields
\meno
$$\Phi(Z) = \varphi(z)+\theta\o\psi(z)
             +\o\theta\psi(z)+\theta\o\theta\chi(z)     $$
\meno
\meno
\hskip 1cm (i) $Q_i+Q_j=Q_k$
\meno
$$\eqalignno{
C_{\varphi_i\varphi_j}^{\varphi_k} &= 2C_{ij}^k   &\cr
C_{\varphi_i\varphi_j}^{\chi'_k} &= C_{ij}^k
{\ts {2\Dt_{ikj}(Q_k/2-\alpha\Dt_k)\over
(2\Dt_k + 1)((Q_k/2)^2 - \Dt_k)} }  &\cr
C_{\o\psi_i\varphi_j}^{\o\psi_k} &= C_{ij}^k
{\ts  {\Dt_{ikj}(\alpha + 1) \over(Q_k/2 + \Dt_k)}}&\cr
C_{\psi_i\varphi_j}^{\psi_k} &= C_{ij}^k
{\ts  {\Dt_{ikj}(\alpha - 1) \over(Q_k/2 - \Dt_k)}}&\cr
C_{\chi'_i\varphi_j}^{\varphi_k} &= C_{ij}^k
 {\ts \Dt_{ikj}\left(\alpha-{Q_i/2\over\Dt_i}\right)} &\cr
C_{\chi'_i\varphi_j}^{\chi'_k} &= -C_{ij}^k  {\ts
     { 2\over (2\Dt_k+1)((Q_k/2)^2-\Dt_k^2)}{\Dt_{ikj}+1\choose 2}
     \left[{Q_k\over 2}({Q_i/2\over\Dt_i}-\alpha)+\Dt_k(1-\alpha
     {Q_i/2\over \Dt_i})\right] } &\cr
C_{\varphi_i\o\psi_j}^{\o\psi'_k} &= C_{ij}^k (-1)^{(2\Dt_i)}{\ts
        \left[2-\Dt_{ikj}{(1+\alpha)\over (Q_k/2-\Dt_k)}\right] }&\cr
C_{\o\psi_i\o\psi_j}^{\varphi_k} &=C_{\o\psi_i\o\psi_j}^{\chi'_k} =0&\cr
C_{\psi_i\o\psi_j}^{\varphi_k} &= C_{ij}^k (-1)^{(2\Dt_i+1)}
  {\ts [\Dt_{ijk}+\Dt_{ikj}\alpha - Q_i ]} &\cr
C_{\psi_i\o\psi_j}^{\chi'_k} &= C_{ij}^k (-1)^{(2\Dt_i+1)}
 {\ts   { \Dt_{ikj}\over (2\Dt_k+1)}{[(Q_k/2 - \alpha\Dt_k)
   (\Dt_{ijk}- Q_i) -\Dt_{ikj}(\Dt_k-\alpha Q_k/2)+2\Dt_k
     (1-\alpha)(Q_k/2+\Dt_k)]\over ((Q_k/2)^2-\Dt_k^2) } } &\cr
C_{\chi'_i\o\psi_j}^{\o\psi_k} &=C_{ij}^k (-1)^{(2\Dt_k)}
 {\ts   \Dt_{ikj}\left[{\Dt_{ijk}(1+\alpha)\over 2(Q_k/2+\Dt_k)}
     \left(1-{Q_i/2\over \Dt_i}\right)+
     \left(\alpha-{Q_i/2\over \Dt_i}\right)\right] } &\cr
C_{\varphi_i\psi_j}^{\psi_k} &=C_{ij}^k (-1)^{(2\Dt_k)}
  {\ts   \left[2+\Dt_{ikj}{(1-\alpha)\over (Q_k/2-\Dt_k)}\right]}&\cr
C_{\o\psi_i\psi_j}^{\varphi_k} &= C_{ij}^k (-1)^{(2\Dt_i+1)}
  {\ts [\Dt_{ijk}-\Dt_{ikj}\alpha + Q_i ]} &\cr
C_{\o\psi_i\psi_j}^{\chi'_k} &= C_{ij}^k (-1)^{(2\Dt_i+1)}
 {\ts   { \Dt_{ikj}\over (2\Dt_k+1)}{[(Q_k/2 - \alpha\Dt_k)
   (\Dt_{ijk}+ Q_i) +\Dt_{ikj}(\Dt_k-\alpha Q_k/2)+2\Dt_k
     (1+\alpha)(Q_k/2-\Dt-k)]\over ((Q_k/2)^2-\Dt_k^2) } } &\cr
C_{\psi_i\psi_j}^{\varphi_k} &=C_{\psi_i\psi_j}^{\chi'_k} =0&\cr
C_{\chi'_i\psi_j}^{\psi_k} &=C_{ij}^k (-1)^{(2\Dt_k)}
 {\ts   \Dt_{ikj}\left[{\Dt_{ijk}(1-\alpha)\over 2(Q_k/2-\Dt_k)}
     \left(1+{Q_i/2\over \Dt_i}\right)+
     \left(\alpha-{Q_i/2\over \Dt_i}\right)\right] } &\cr
C_{\varphi_i\chi'_j}^{\varphi_k} &= C_{ij}^k
  {\ts \left[\Dt_{ikj}\alpha-2\Dt_i-\Dt_{ijk}{Q_j/2\over \Dt_j}\right]} &\cr
C_{\varphi_i\chi'_j}^{\chi'_k} &= C_{ij}^k {\ts\biggl\{
 2+ {\Dt_{ikj}\over (2\Dt_k+1)}{1\over ((Q_k/2)^2 - \Dt_k^2)} \biggl[
 2(2\Dt_k+1)(\Dt_k-\alpha Q_k/2)-(\Dt_{ikj}+ 1)(\Dt_k-\alpha Q_k/2)}&\cr
&\quad\quad {\ts - (Q_k/2-\alpha\Dt_k)\left( Q_i+(\Dt_{ijk}-1)
{Q_j/2\over \Dt_j}
        \right)\biggr]\biggr\} } &\cr
C_{\o\psi_i\chi'_j}^{\o\psi_k} &=C_{ij}^k
{\ts     \left[\Dt_{ikj}\alpha -(\Dt_{ijk}+ Q_i)+\Dt_{ikj}{\Dt_{ijk}
   (1+\alpha)\over 2(Q_k/2-\Dt_k)}\left( 1-{Q_j/2\over\Dt_j}\right)
   \right]} &\cr
C_{\psi_i\chi'_j}^{\psi_k} &=C_{ij}^k
 {\ts    \left[\Dt_{ikj}\alpha +(\Dt_{ijk}- Q_i)+\Dt_{ikj}{\Dt_{ijk}
   (1-\alpha)\over 2(Q_k/2-\Dt_k)}\left( 1+{Q_j/2\over\Dt_j}\right)
   \right]} &\cr
C_{\chi'_i\chi'_j}^{\varphi_k} &=C_{ij}^k \biggl[
 {\ts  {\Dt_{ijk}+1\choose 2}\left(1+{Q_i Q_j\over 4\Dt_i\Dt_j}\right)
    +{Q_i(\Dt_{ijk}+1)\over 4\Dt_i}\left(\Dt_{ikj}\alpha-Q_i-\Dt_{ijk}
    {Q_j/2\over \Dt_j}\right) }  &\cr
&\quad\quad {\ts - {Q_j(\Dt_{ijk}+1)\over 4\Dt_j}\Dt_{ikj}\left(\alpha
    -{\Dt_i/2\over \Dt_i}\right)\biggr] } &\cr
C_{\chi'_i\chi'_j}^{\chi'_k} &=C_{ij}^k
 {\ts\left[  {\Dt_{ijk}\choose 2}{\Dt_{ikj}(Q_k/2-\alpha\Dt_k)\over
   (2\Dt_k+1)((Q_k/2)^2-\Dt_k^2)}+\Dt_{ijk}\Dt_{ikj}
{(Q_k/2-\alpha\Dt_k)\over ((Q_k/2)^2-\Dt_k^2)}+\Dt_{ikj}\alpha-Q_i\right]}
&\cr
  &\quad\quad {\ts +{Q_i\Dt_{ijk}\over 4\Dt_i}
     C_{\varphi_i\chi'_j}^{\chi'_k}
   -{Q_j\Dt_{ijk}\over 4\Dt_j}C_{\chi'_i\varphi_j}^{\chi'_k}
   +{Q_i Q_j\over 8\Dt_i\Dt_j}{\Dt_{ijk}\choose 2}
       C_{\varphi_i\varphi_j}^{\chi'_k}  }&\cr }$$
\bigno
\hskip 1cm (ii) $Q_i+Q_j=Q_k-1$
\meno
$$\eqalignno{
C_{\varphi_i\varphi_j}^{\o\psi_k} &=C_{ij}^k{\ts {1\over (Q_k/2+\Dt_k)}}&\cr
C_{\o\psi_i\varphi_j}^{\varphi_k} &= C_{\o\psi_i\varphi_j}^{\chi'_k} =0 &\cr
C_{\psi_i\varphi_j}^{\varphi_k} &= C_{ij}^k &\cr
C_{\psi_i\varphi_j}^{\chi'_k} &=C_{ij}^k{\ts{(\Dt_{ikj}+1/2)\over (2\Dt_k+1)
                (Q_k/2+\Dt_k) }}&\cr
C_{\chi'_i\varphi_j}^{\o\psi_k} &=-C_{ij}^k{\ts{(\Dt_{ikj}+1/2)\over
               2(Q_k/2+\Dt_k) } \left(1+{Q_i/2\over \Dt_i}\right)}&\cr
C_{\varphi_i\o\psi_j}^{\varphi_k} &= C_{\varphi_i\o\psi_j}^{\chi'_k} =0 &\cr
C_{\o\psi_i\o\psi_j}^{\o\psi_k} &= C_{\o\psi_i\o\psi_j}^{\psi_k} =0 &\cr
C_{\psi_i\o\psi_j}^{\o\psi_k} &=C_{ij}^k (-1)^{(2\Dt_i+1)}
      {\ts {(Q_j/2+\Dt_j)\over (Q_k/2+\Dt_k)}}&\cr
C_{\chi'_i\o\psi_j}^{\varphi_k} &= C_{\chi'_i\o\psi_j}^{\chi'_k} =0 &\cr
C_{\varphi_i\psi_j}^{\varphi_k} &= C_{ij}^k (-1)^{(2\Dt_i+1)} &\cr
C_{\varphi_i\psi_j}^{\chi'_k} &= C_{ij}^k (-1)^{(2\Dt_i)}
            {\ts {(\Dt_{jki}+1/2)\over (2\Dt_k+1)
                (Q_k/2+\Dt_k) }}&\cr
C_{\o\psi_i\psi_j}^{\o\psi_k} &=C_{ij}^k (-1)^{(2\Dt_i+1)}
      {\ts {(Q_i/2+\Dt_i)\over (Q_k/2+\Dt_k)}}&\cr
C_{\psi_i\psi_j}^{\psi_k} &= C_{ij}^k (-1)^{(2\Dt_i+1)} &\cr
C_{\chi'_i\psi_j}^{\varphi_k} &=C_{ij}^k (-1)^{(2\Dt_i+1)}
 {\ts {(\Dt_{ijk}+1/2)\over 2} \left(1+{Q_i/2\over \Dt_i}\right)}&\cr
C_{\chi'_i\psi_j}^{\chi'_k} &=C_{ij}^k (-1)^{(2\Dt_i+1)}
 {\ts {\Dt_{ikj}+1/2\over 2}{(\Dt_i+\Dt_j+\Dt_k+1/2)\over (2\Dt_k+1)
   (Q_k/2+\Dt_k)} \left(1+{Q_i/2\over \Dt_i}\right)}&\cr
C_{\varphi_i\chi'_j}^{\o\psi_k} &=C_{ij}^k
 {\ts \left[ 1-{(\Dt_{ikj}+1/2+Q_i)\over  2(Q_k/2+\Dt_k)}
 {(\Dt_{ijk}-1/2)\over  (Q_k/2+\Dt_k)} {Q_j\over 4\Dt_j} \right] }&\cr
C_{\o\psi_i\chi'_j}^{\varphi_k} &= C_{\o\psi_i\chi'_j}^{\chi'_k} =0 &\cr
C_{\psi_i\chi'_j}^{\varphi_k} &=-C_{ij}^k
 {\ts {(\Dt_{ijk}+1/2)\over 2} \left(1+{Q_j/2\over \Dt_j}\right)}&\cr
C_{\psi_i\chi'_j}^{\chi'_k} &=C_{ij}^k
 {\ts {\Dt_{jki}+1/2\over 2}{(\Dt_i+\Dt_j+\Dt_k+1/2)\over (2\Dt_k+1)
   (Q_k/2+\Dt_k)} \left(1+{Q_j/2\over \Dt_j}\right)}&\cr
C_{\chi'_i\chi'_j}^{\o\psi_k} &=C_{ij}^k
 {\ts \left[ {1\over 2(Q_k/2+\Dt_k)}{\Dt_{ijk}+1/2\choose 2}+
   {(\Dt_{ijk}+1/2)\over 2}\right]
+{Q_i(\Dt_{ijk}+1/2)\over 4\Dt_i}C_{\varphi_i\chi'_j}^{\o\psi_k} }&\cr
  &\quad\quad {\ts
   -{Q_j(\Dt_{ijk}+1/2)\over 4\Dt_j}C_{\chi'_i\varphi_j}^{\o\psi_k}
   +{Q_i Q_j\over 8\Dt_i\Dt_j}{\Dt_{ijk}+1/2\choose 2}
       C_{\varphi_i\varphi_j}^{\o\psi_k}  }&\cr }$$
\bigno
\hskip 1cm (iii) $Q_i+Q_j=Q_k+1$
\meno
$$\eqalignno{
C_{\varphi_i\varphi_j}^{\psi_k} &=C_{ij}^k{\ts {1\over (Q_k/2-\Dt_k)}}&\cr
C_{\o\psi_i\varphi_j}^{\varphi_k} &= C_{ij}^k &\cr
C_{\o\psi_i\varphi_j}^{\chi'_k} &=C_{ij}^k{\ts{(\Dt_{ikj}+1/2)\over (2\Dt_k+1)
                (Q_k/2-\Dt_k) }}&\cr
C_{\psi_i\varphi_j}^{\varphi_k} &= C_{\psi_i\varphi_j}^{\chi'_k} =0 &\cr
C_{\chi'_i\varphi_j}^{\psi_k} &=-C_{ij}^k{\ts{(\Dt_{ikj}+1/2)\over
               2(Q_k/2-\Dt_k) } \left(1-{Q_i/2\over \Dt_i}\right)}&\cr
C_{\varphi_i\o\psi_j}^{\varphi_k} &= C_{ij}^k (-1)^{(2\Dt_i+1)} &\cr
C_{\varphi_i\o\psi_j}^{\chi'_k} &= C_{ij}^k (-1)^{(2\Dt_i)}
            {\ts {(\Dt_{jki}+1/2)\over (2\Dt_k+1)
                (Q_k/2-\Dt_k) }}&\cr
C_{\o\psi_i\o\psi_j}^{\o\psi_k} &= C_{ij}^k (-1)^{(2\Dt_i+1)} &\cr
C_{\psi_i\o\psi_j}^{\psi_k} &=C_{ij}^k (-1)^{(2\Dt_i+1)}
      {\ts {(Q_i/2-\Dt_i)\over (Q_k/2-\Dt_k)}}&\cr
C_{\chi'_i\o\psi_j}^{\varphi_k} &=C_{ij}^k (-1)^{(2\Dt_i)}
 {\ts {(\Dt_{ijk}+1/2)\over 2} \left(1-{Q_i/2\over \Dt_i}\right)}&\cr
C_{\chi'_i\o\psi_j}^{\chi'_k} &=C_{ij}^k (-1)^{(2\Dt_i)}
 {\ts {\Dt_{ikj}+1/2\over 2}{(\Dt_i+\Dt_j+\Dt_k+1/2)\over (2\Dt_k+1)
   (Q_k/2-\Dt_k)} \left(1-{Q_i/2\over \Dt_i}\right)}&\cr
C_{\varphi_i\psi_j}^{\varphi_k} &= C_{\varphi_i\psi_j}^{\chi'_k} =0 &\cr
C_{\o\psi_i\psi_j}^{\psi_k} &=C_{ij}^k (-1)^{(2\Dt_i+1)}
      {\ts {(Q_j/2-\Dt_j)\over (Q_k/2-\Dt_k)}}&\cr
C_{\psi_i\psi_j}^{\o\psi_k} &= C_{\psi_i\psi_j}^{\psi_k} =0 &\cr
C_{\chi'_i\psi_j}^{\varphi_k} &= C_{\chi'_i\psi_j}^{\chi'_k} =0 &\cr
C_{\varphi_i\chi'_j}^{\o\psi_k} &=C_{ij}^k
 {\ts \left[ -1-{(\Dt_{ikj}+1/2-Q_i)\over  2(Q_k/2-\Dt_k)}
 +{(\Dt_{ijk}-1/2)\over  (Q_k/2-\Dt_k)} {Q_j\over 4\Dt_j} \right] }&\cr
C_{\o\psi_i\chi'_j}^{\varphi_k} &=-C_{ij}^k
 {\ts {(\Dt_{ijk}+1/2)\over 2} \left(1-{Q_j/2\over \Dt_j}\right)}&\cr
C_{\o\psi_i\chi'_j}^{\chi'_k} &=C_{ij}^k
 {\ts -{\Dt_{jki}+1/2\over 2}{(\Dt_i+\Dt_j+\Dt_k+1/2)\over (2\Dt_k+1)
   (Q_k/2-\Dt_k)} \left(1-{Q_j/2\over \Dt_j}\right)}&\cr
C_{\psi_i\chi'_j}^{\varphi_k} &= C_{\psi_i\chi'_j}^{\chi'_k} =0 &\cr
C_{\chi'_i\chi'_j}^{\o\psi_k} &=C_{ij}^k
 {\ts \left[ -{1\over 2(Q_k/2-\Dt_k)}{\Dt_{ijk}+1/2\choose 2}+
   {(\Dt_{ijk}+1/2)\over 2}\right]
+{Q_i(\Dt_{ijk}+1/2)\over 4\Dt_i}C_{\varphi_i\chi'_j}^{\psi_k} }&\cr
  &\quad\quad {\ts
   -{Q_j(\Dt_{ijk}+1/2)\over 4\Dt_j}C_{\chi'_i\varphi_j}^{\psi_k}
   +{Q_i Q_j\over 8\Dt_i\Dt_j}{\Dt_{ijk}+1/2\choose 2}
       C_{\varphi_i\varphi_j}^{\psi_k}  }&\cr }$$
\bigno
\bigno
{\cl{\bf Appendix B}}
\bigno
In this appendix we present the component expansion of the normal ordered
products $N_s(..)$.
The four types of fields occuring in the regular part of the OPE
$\Phi_i(Z_1)\Phi_j(Z_2)$ are
\meno
$$ N_s(\Phi_j\partial^n\Phi_i)=\cases{ N(\varphi_j\partial^n\varphi_i) \crns
    +\theta[\>(-1)^{2\Dt_i} N(\o\psi_j\partial^n\varphi_i)+
      N(\varphi_j\partial^n\o\psi_i)\>] \crns
    +\o\theta[\>(-1)^{2\Dt_i} N(\psi_j\partial^n\varphi_i)+
      N(\varphi_j\partial^n\psi_i)\>] \crns
    +\theta\o\theta[\>N(\chi_j\partial^n\varphi_i)+
      N(\varphi_j\partial^n\chi_i)-
    (-1)^{2\Dt_i} N(\psi_j\partial^n\o\psi_i) \cr
    \quad\quad +(-1)^{2\Dt_i} N(\o\psi_j\partial^n\psi_i)\>] \cr } $$
\meno
$$ N_s(\Phi_j D\partial^n\Phi_i)=\cases{ N(\varphi_j\partial^n\o\psi_i) \crns
    +\theta (-1)^{2\Dt_i+1} N(\o\psi_j\partial^n\o\psi_i) \crns
    +\o\theta[\>(-1)^{2\Dt_i+1} N(\psi_j\partial^n\o\psi_i)+
      N(\varphi_j\partial^n\chi_i)+{\ts {1\over 2}}
      N(\varphi_j\partial^{n+1}\varphi_i)\>] \crns
    +\theta\o\theta[\>N(\chi_j\partial^n\o\psi_i)-
    (-1)^{2\Dt_i} N(\o\psi_j\partial^n\chi_i)-
      {\ts {1\over 2}} N(\varphi_j\partial^{n+1}\o\psi_i) \cr
     \quad\quad-{\ts {1\over 2}}
    (-1)^{2\Dt_i} N(\o\psi_j\partial^{n+1}\varphi_i)\>]\cr}$$
\meno
$$ N_s(\Phi_j \o{D}\partial^n\Phi_i)=\cases{ N(\varphi_j\partial^n\psi_i)
\crns
    +\theta[\>(-1)^{2\Dt_i+1} N(\o\psi_j\partial^n\psi_i)-
      N(\varphi_j\partial^n\chi_i)+{\ts {1\over 2}}
      N(\varphi_j\partial^{n+1}\varphi_i)\>] \crns
    +\o\theta (-1)^{2\Dt_i+1} N(\psi_j\partial^n\psi_i) \crns
    +\theta\o\theta[\>N(\chi_j\partial^n\psi_i)-
    (-1)^{2\Dt_i} N(\psi_j\partial^n\chi_i)+
      {\ts {1\over 2}} N(\varphi_j\partial^{n+1}\psi_i)\cr
     \quad\quad +{\ts {1\over 2}}
    (-1)^{2\Dt_i} N(\psi_j\partial^{n+1}\varphi_i)\>]\cr}$$
\meno
$$ N_s(\Phi_j[\o{D},D]\partial^n\Phi_i)=\cases{ 2N(\varphi_j\partial^n\chi_i)
 \crns
    +\theta[\>(-1)^{2\Dt_i} 2N(\o\psi_j\partial^n\chi_i)+
      N(\varphi_j\partial^{n+1}\o\psi_i)\>] \crns
    +\o\theta[\>(-1)^{2\Dt_i} 2N(\psi_j\partial^n\chi_i)-
      N(\varphi_j\partial^{n+1}\varphi_i)\>] \crns
    +\theta\o\theta[\>2N(\chi_j\partial^n\chi_i)+
     {\ts {1\over 2}}  N(\varphi_j\partial^{n+2}\varphi_i)-
    (-1)^{2\Dt_i} 2N(\o\psi_j\partial^{n+1}\psi_i) \cr
   \quad\quad- (-1)^{2\Dt_i} 2N(\psi_j\partial^{n+1}\o\psi_i)\>] \cr } $$
\bigno
\bigno
\bigno
\cl{\bf References}
\medskip
\halign{\hfil $ # $ \quad & \rm # \hfill \cr
\q{1}&  D. Gepner,  Nucl. Phys. B{\bf 296}
  (1988) 757 \crns
\q{2}&  D. Gepner,  Phys. Lett. {\bf 199}B
  (1987) 380 \crns
\q{3}&  D. Gepner,  Nucl. Phys. B{\bf 311}
  (1988) 191 \crns
\q{4}&  T. Banks, L.J. Dixon, D. Friedan, E. Martinec,
       Nucl. Phys. B{\bf 299}   (1988) 613 \crns
\q{5}&  W. Boucher, D. Friedan, A. Kent,
  Phys. Lett. {\bf 172}B (1986) 316 \crns
\q{6}&  Y. Kazama, H. Suzuki,
       Nucl. Phys. B{\bf 321}   (1989) 232 \crns
\q{7}&  Y. Kazama, H. Suzuki,
       Mod. Phys. Lett. A4 (1989) 235 \crns
\q{8}&  A.B. Zamolodchikov,  Theor. Math. Phys. {\bf 65} (1986) 1205  \crns
\q{9}&  R. Blumenhagen, M. Flohr, A. Kliem, W. Nahm, A. Recknagel,
R. Varnhagen, \cr
     &  Nucl. Phys. B{\bf 361}   (1991) 255 \crns
\q{10}&  H.G. Kausch, G.M.T. Watts,
       Nucl. Phys. B{\bf 354}   (1991) 740 \crns
\q{11}&  J.M. Figueroa-O'Farrill, S. Schrans,
  Phys. Lett. {\bf 245}B (1990) 471 \crns
\q{12}&  R. Blumenhagen, W. Eholzer, A. Honecker, R. H\"ubel, \cr
      & preprint BONN-HE-92-02  to appear in Int. J. Mod. Phys. A\crns
\q{13}&  T. Inami, Y. Matsuo, I. Yamanaka, Int. J. Mod. Phys.
          A5 (1990) 4441 \crns
\q{14}&  L.J. Romans, Nucl. Phys. B{\bf 369}   (1992) 403 \crns
\q{15}&  R. Blumenhagen,
       preprint BONN-HE-91-20  to appear in Nucl. Phys. B\crns
\q{16}&  P. di Vecchia, J.L. Petersen, H.B. Zheng
     Phys. Lett. {\bf 162}B (1985) 327 \crns
\q{17}&  P. West, 'Introduction to supersymmetry and supergravity',
          second edition 1990 \cr &  World Scientific, Singapore\crns
\q{18}&  G. Mussardo, G. Sotkov, M. Stanishkov, Int. J. Mod. Phys.
          A4 (1989) 1135 \crns
\q{19}&  W. Lerche, C. Vafa, N.P. Warner,
       Nucl. Phys. B{\bf 324}  (1989) 427 \crns
\q{20}&  S. Odake,
       Mod. Phys. Lett. A4 (1989) 557 \crns
\q{21}&  J.L. Cardy,
       Nucl. Phys. B{\bf 370}  (1986) 186 \crns
\q{22}&   W. Eholzer, M. Flohr, A. Honecker, R. H\"ubel, W. Nahm, R.
             Varnhagen,\cr
      & preprint BONN-HE-91-22  to appear in Nucl. Phys. B\crns
\q{23}&  A.N. Schellekens, S. Yankielowicz,
       Nucl. Phys. B{\bf 327}  (1990) 673 \crns
\q{24}&  F. Ravanini, S.K. Yang,
  Phys. Lett. {\bf 195}B (1990) 202 \crns
\q{25}&  M. Flohr,
       preprint BONN-HE-92-08   \crns
\q{26}&  A. Schwimmer, N. Seiberg,
  Phys. Lett. {\bf 184}B (1987) 191 \crns
\q{27}&  S. Odake,
       Int. J. Mod. Phys.  A5 (1990) 897 \crns
\q{28}&  S. Odake,
       Mod. Phys. Lett. A5 (1990) 561 \crns
\q{29}&  L.J. Dixon, M.E. Peskin, J. Lykken,
       Nucl. Phys. B{\bf 325}  (1989) 329 \crns
\q{30}&  L. Bars, Nucl. Phys. B{\bf 334}  (1990) 125 \crns  }
\bye
\end